\documentclass[notitlepage,superscriptaddress,pre,longbibliography]{revtex4-1}%
\usepackage{placeins}
\usepackage{pbox}
\usepackage{color}
\usepackage{relsize}
\usepackage{graphicx}
\usepackage[intlimits]{amsmath}
\usepackage{amsxtra, amssymb,fancyhdr, amsthm,latexsym}
\usepackage{MathShorthand}
\usepackage{cases}
\usepackage{verbatim}
\usepackage{float}
\usepackage[hidelinks]{hyperref}
\usepackage{textcomp}
\usepackage{epstopdf}
\usepackage[margin=0.75in]{geometry}
\usepackage{amsfonts}
\usepackage{amssymb}%
\providecommand{\U}[1]{\protect\rule{.1in}{.1in}}
\newtheorem{theorem}{Theorem}[section]

\newtheorem{lem}[theorem]{Lemma}

\newtheorem{thm}[theorem]{Theorem}

\newtheorem{prop}[theorem]{Proposition}
\newtheorem{conjecture}[theorem]{Conjecture}

\begin{document}
\begin{empty}
\title{Toroidal Vortex Filament Knots \& Links: Existence, Stability
and Dynamics}
\author{T. Kolokolnikov}
\email{tkolokol@gmail.com}
\affiliation{Department of Mathematics and Statistics,
Dalhousie University Halifax,
Nova Scotia, B3H3J5, Canada}
\author{Christopher Ticknor}
\email{cticknor@lanl.gov}
\affiliation{Theoretical Division, Los Alamos National Laboratory, Los Alamos, New Mexico 87545, USA}
\author{P.~G. Kevrekidis}
\email{kevrekid@umass.edu}
\affiliation{Department of Mathematics and Statistics, University of Massachusetts,
Amherst, Massachusetts 01003-4515 USA}
\date{\today}
\begin{abstract}
Using the Klein-Majda-Damodaran model of nearly-parallel vortex filaments, we
construct vortex knots and links on a torus involving periodic boundary
conditions and analyze their stability. For a special
class of vortex knots -- toroidal knots -- we give a full characterization of
both their energetic and dynamical stability. In addition to providing
explicit expressions for the relevant waveforms, we derive explicit
formulas  for their
stability boundaries. These include simple
links and\ different realizations of a trefoil knot.
It is shown that a ring
of more than 7 filaments can potentially be stablized by giving it a slight twist and
connecting neighbouring filaments on a torus.
In addition to rings, (helical) filament lattice configurations are
also considered and are found to be dynamically stable for all rotation
frequencies and also energetically stable for sufficiently fast rotations.
Numerical simulations are used
to compare the Klein-Majda-Damodaran model with  the full
three-dimensional
(3D)
Gross-Pitaevskii\ equations
as well as to confirm the analytical theory. Potential differences between the
quasi-one-dimensional and the fully 3D description  are also discussed.
\end{abstract}
\maketitle
\end{empty}

\section{Introduction}

States bearing topological charge constitute a principal theme within a
variety of areas in Physics, including (but not limited to)
optics~\cite{Pismen}, condensed matter~\cite{Donnelly,Blatter1994}, as well as
hydrodynamics~\cite{Saffman}. The experimental realization of Bose-Einstein
condensates (BECs) has offered a platform where numerous vortical excitations
could be explored, and their interactions with each other and with external
potential landscapes could be monitored in a systematic, time-resolved
manner~\cite{becbook1,becbook2,siambook}. This has led to a significantly
enhanced understanding of the role of vortical patterns in BEC dynamics, as
well as in quantum turbulence, which has by now been summarized in a
substantial number of review
publications~\cite{fetter2,Alexander2001,mplb04,Komineas2007,White2013,TSATSOS20161}%
.

The gradual formulation of an understanding of the building blocks such as
vortex lines and vortex rings as summarized in the above studies has propelled
a considerable volume of ongoing interest towards the formulation, dynamical
monitoring and qualitative understanding of more elaborate structures such as
vortex knots and links. These have been explored chiefly in a homogeneous
(density background) setting as, e.g.,
in~\cite{RSB1999,pre09,pre12,nature13,Proment_2014,brachet16,nature16,Ruban2018a,Ruban2018b}%
. Recent work~\cite{rubanus} has argued that upon suitable (anisotropic)
trapping conditions such knot structures may be long-lived, while experiments
with spinor BECs have spearheaded the realization/observation of the
structures~\cite{nature16b,Leeeaao3820}. It is worth noting, in passing, that
such structures are not only relevant in BECs but in numerous other areas
including, e.g., nonlinear optics~\cite{scirep12}, but also DNA
strands~\cite{Shimokawa20906}, magnetic fields in plasmas~\cite{nature13b},
classical fluids \cite{lim1997role}, superfluids \cite{caplan2014scattering,
baggaley2011spectrum}; and helical filaments in the wake of turbines
\cite{okulov2010maximum, okulov2004stability, quaranta2015long,
leweke2014long}.

Our starting point in the present work will be rather different than that of
most of the above studies. We will start from an effective
quasi-one-dimensional mathematical description of vortex filaments developed
in~\cite{klein1995simplified} (see also~\cite{lions2000equilibrium} for an
equilibrium statistical theory and~\cite{contreras2016nearly} for a recent
dynamical analysis of the model). We will use the latter as a framework for
obtaining exact analytical solutions for co-rotating (helical) vortex
filaments. The latter through the use of periodic boundary conditions will
formulate structures akin to vortical knots and links. Upon identifying such
states and parametrizing them by a pair of integer indices, we will analyze
their existence (e.g. frequency and radius of rotation in section 2), as well
as stability (in section 3) properties. In section 4, we examine a lattice of
such filaments. In section 5, we return to the original motivating problem of
the 3-dimensional prototypical model of BECs (the Gross-Pitaevskii (GP)
equation~\cite{becbook1,becbook2,siambook}) and explore the validity of our
existence and stability conclusions therein. Finally, we summarize our
findings and present our conclusions, as well as some possible directions of
future study in section 6.

\section{Existence of Vortex Filament States}


In \cite{klein1995simplified, lions2000equilibrium} the authors derived a
simplified model describing the evolution of $K$ nearly-parallel vortex
filaments -- the so-called Klein--Majda--Damodaran (KMD) model. The reduced
equations they derived are%
\begin{equation}
-i\frac{\partial}{\partial t}X_{k}=D\frac{\partial^{2}}{\partial z^{2}}%
X_{k}+\sum_{j\neq k}\frac{X_{k}-X_{j}}{\left\vert X_{k}-X_{j}\right\vert ^{2}%
},\ \ \ k=1\ldots K.\label{reduced}%
\end{equation}

Here, $z$ denotes the direction that is nearly parallel to all the filaments;
and $X_{k}(z,t)\in%
\mathbb{R}
^{2}$ is the two-dimensional position of the $k-$th (topologically charged)
filament at height $z$ and time $t$.

The goal of this paper is to study the stability of helical \textquotedblleft
co-rotating\textquotedblright\ vortex filaments using the reduced equations
(\ref{reduced}). We also, however, endeavor to return to the original
Gross-Pitaevskii model from which this reduced dynamics is obtained and to
compare the predictions of the effective model with the original one. We
assume that each filament is rotating with the same angular velocity $\Omega$
without changes in shape, so that the whole configuration undergoes a
\textquotedblleft rigid\textquotedblright\ rotation. These are some of the
simplest nontrivial filament configurations. By analogy to point vortex
literature, we refer to these configurations as \emph{relative equilibria}
\cite{barry2012relative, newton2001n, aref2003vortex, palmore1982relative},
i.e., equilibria in the rotational frame of reference. Assuming that the
system rotates with a rate $\Omega,$ we make a change of variables%
\begin{equation}
X_{k}(z,t)=e^{i\Omega t}\xi_{k}(z,t)
\end{equation}
so that the $\xi_{k}$ satisfy%
\begin{equation}
-i\frac{\partial}{\partial t}\xi_{k}=D\frac{\partial^{2}}{\partial z^{2}}%
\xi_{k}-\Omega\xi_{k}+\sum_{j\neq k}\frac{\xi_{k}-\xi_{j}}{\left\vert \xi
_{k}-\xi_{j}\right\vert ^{2}}. \label{dyn}%
\end{equation}

Relative equilibria are time-independent steady states of this system. More
generally, we also examine what happens when adding \textquotedblleft
relaxation\textquotedblright\ as follows:%
\begin{equation}
\left(  -i\gamma_{1}+\gamma_{2}\right)  \frac{\partial}{\partial t}\xi
_{k}=D\frac{\partial^{2}}{\partial z^{2}}\xi_{k}-\Omega\xi_{k}+\sum_{j\neq
k}\frac{\xi_{k}-\xi_{j}}{\left\vert \xi_{k}-\xi_{j}\right\vert ^{2}}.
\label{general}%
\end{equation}
The latter model has been used in the context of vortices in order to study
their dynamics in finite temperature settings~\cite{DGPE,zueva} and relevant
ideas have even been extended to recent experiments measuring, e.g., the rate
of vortex spiraling out of the condensate in connection with the relaxational
term in the corresponding dynamics~\cite{yongil}. The limit $\gamma
_{1}\rightarrow0$ (or $\gamma_{2}\rightarrow\infty)$ can be thought of as an
overdamped limit. After time-rescaling, the overdamped system leads to
\begin{equation}
\frac{\partial}{\partial t}\xi_{k}=D\frac{\partial^{2}}{\partial z^{2}}\xi
_{k}-\Omega\xi_{k}+\sum_{j\neq k}\frac{\xi_{k}-\xi_{j}}{\left\vert \xi_{k}%
-\xi_{j}\right\vert ^{2}}. \label{ener}%
\end{equation}

The equilibrium (time-independent)\ confirgurations are solutions of the
system (\ref{general}); at the same time they are are also steady states for
both (\ref{dyn})\ and (\ref{ener}). However it turns out that the stability of
the two systems (\ref{dyn})\ and (\ref{ener})\ can be different. We shall
refer to stability with respect to (\ref{ener})\ as \textbf{energetic
stability; }whereas the stability with respect to (\ref{dyn})\ will be called
\textbf{dynamical stability}. This nomenclature is justified by thinking of
(\ref{ener})\ as the gradient flow of the associated energy functional
discussed, e.g., in~\cite{klein1995simplified,contreras2016nearly}. We will
show that there are \emph{stable} equilibrium solutions of (\ref{dyn})\ which
are unstable with respect to the system~(\ref{ener}). This is a feature that
is common in Hamiltonian systems in connection to their dissipative
counterparts; see, e.g.,~\cite{ruste} for a relevant discussion of dynamical
and energetic stability.

The questions that we ask are the following:

\begin{itemize}
\item What are the steady states of (\ref{dyn})?

\item What is their dynamical stability (i.e. stability with respect to
(\ref{dyn}))?

\item What is their energetic stability (i.e. stability with respect to
(\ref{ener}))?
\end{itemize}

The stability is intimately connected with the selection of boundary
conditions. In this work we assume the following \emph{doubly-periodic
boundary conditions}%
\begin{equation}
z\in\left[  0,P\right]  ;\ \ \ \ \ \ \xi_{k}(P,t)=\xi_{k+q}\left(  0,t\right)
\ \ \ \ (\operatorname{mod}K). \label{BC}%
\end{equation}
The \textquotedblleft usual\textquotedblright\ periodic BC correspond to the
case where $q$ is an integer multiple of $K.$ When visualized on a torus, such
doubly-periodic solutions represent knots and links (with the number of links
being given by the $\gcd\left(  K,q\right)  $). The simplest such knots and
links, corresponding to helical filaments, are the so-called \emph{toroidal}
knots, and are illustrated in Figure \ref{fig:1}; see also Table
\ref{table:ener} below. For example the trefoil knot can be realized as a
$\left(  K,q\right)  =(2,3)$ toroidal knot.

\begin{figure}[tb]
\begin{center}
\includegraphics[width=1\textwidth]{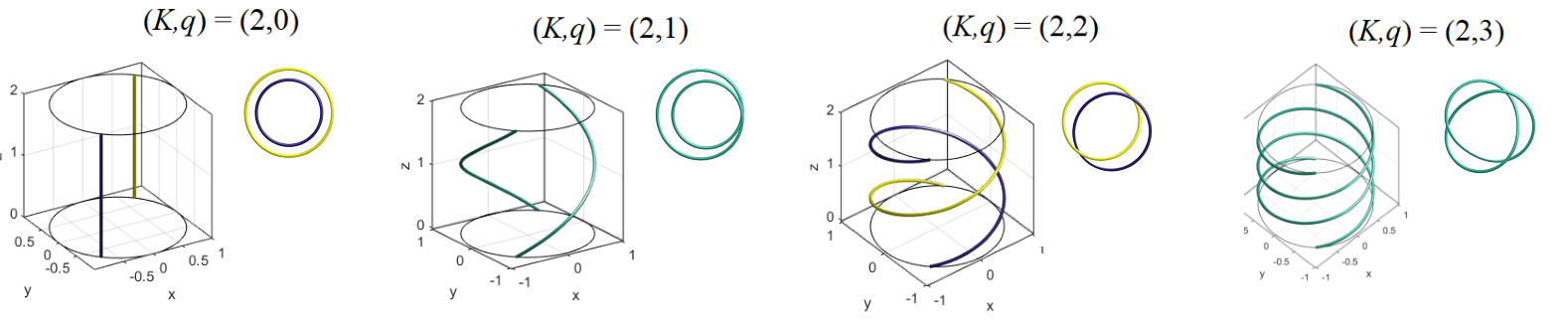}
\end{center}
\caption{Two graphical representations of toroidal vortex filament knots for
several values of $(K,q)$ as indicated. On the left is the cartesian
representation with $(x,y)=\xi_{k}(z)$. On the right is toroidal
representation with $z$ wrapping around the center of the torus. Refer also to
Table \ref{table:ener} }%
\label{fig:1}%
\end{figure}

The simplest configuration consists of straight filaments, where each z
cross-section corresponds to point vortices in a relative equilibrium. The
corresponding dynamics is that of co-rotating 2D vortices, trivially extended
into the third dimension.
As we will show below, these structures are dynamically stable with respect to
the reduced equations (\ref{reduced}), as long as the underlying point vortex
configuration is stable.

A more interesting steady state consists of helical filaments having the form%
\begin{equation}
\xi_{k}(t,z)=e^{zi\omega}\eta_{k} \label{1244}%
\end{equation}
with $\eta_{k}$ satisfiying%
\begin{equation}
0=-\Gamma\eta_{k}+\sum_{j\neq k}\frac{\eta_{k}-\eta_{j}}{\left\vert \eta
_{k}-\eta_{j}\right\vert ^{2}}\ \ \ \text{where\ \ \ }\Gamma=D\omega
^{2}+\Omega. \label{vortex-ss}%
\end{equation}
In other words, $\left\{  \eta_{k}\right\}  ,k=1\ldots K$ are relative
equilibria of the associated point-vortex problem, corresponding to a shifted
frequency $\Gamma=\Omega+D\omega^{2}.$ Conversely, all helical filament states
of the form (\ref{1244})\ correspond to equilibria of the point-vortex problem
(\ref{vortex-ss}). Note also that $\Gamma$ in (\ref{vortex-ss}) can be set to
one by rescaling, so that this problem is parameter-free.

We remark that the solution to (\ref{vortex-ss}) exists only if $\Gamma>0$
\cite{chen2013collective}. To see this, take the dot product of
(\ref{vortex-ss}) with $\eta_{k}$ and sum over $k.$ We then obtain%
\[
\Gamma\sum_{k}\left\vert \eta_{k}\right\vert ^{2}=\sum_{k}\sum_{j\neq k}%
\frac{\left\vert \eta_{k}\right\vert ^{2}-\eta_{j}\cdot\eta_{k}}{\left\vert
\eta_{k}-\eta_{j}\right\vert ^{2}}=\sum_{k}\sum_{j>k}\frac{\left\vert \eta
_{k}\right\vert ^{2}-2\eta_{j}\cdot\eta_{k}+\left\vert \eta_{j}\right\vert
^{2}}{\left\vert \eta_{k}-\eta_{j}\right\vert ^{2}}=\frac{\left(  K-1\right)
K}{2},
\]
so that
\begin{equation}
\Gamma=\frac{\left(  K-1\right)  K}{2}\frac{1}{\sum_{k}\left\vert \eta
_{k}\right\vert ^{2}}>0.\label{644}%
\end{equation}

A natural realization of the boundary conditions (\ref{BC}) is by placing
vortices uniformly along a ring,%
\begin{equation}
\eta_{k}=re^{i2\pi k/K}%
\end{equation}
while setting the frequency $\omega$ in (\ref{1244}) to be
\begin{equation}
\omega=\frac{2\pi}{P}\frac{q}{K},\ \ q\in%
\mathbb{Z}
, \label{omega}%
\end{equation}
so that the doubly-periodic boundary conditions are automatically satisfied.
Then we have%
\[
\sum_{j\neq k}\frac{\eta_{k}-\eta_{j}}{\left\vert \eta_{k}-\eta_{j}\right\vert
^{2}}=\frac{e^{i2\pi k/K}}{r}\frac{\left(  K-1\right)  }{2},
\]
so that $r$ is given by%
\begin{equation}
r^{2}=\frac{K-1}{2\Gamma} \label{7:52c}%
\end{equation}
(as can also be seen from (\ref{644})). This leads to what we shall call a
filament ring state:

\begin{prop}
\label{prop:ss}\textbf{ ((K,q) filament ring).} For any integers $q,K$, there
is a steady state of (\ref{ener})\ that has the form%
\begin{equation}
\xi_{k}(t,z)=re^{zi\omega}e^{i2\pi k/K}%
\end{equation}
where
\begin{equation}
\omega=\frac{2\pi}{P}\frac{q}{K},\ \ r^{2}=\frac{K-1}{2\left(  D\omega
^{2}+\Omega\right)  } \label{7:52d}%
\end{equation}
Such a steady state satisfies the boundary condtions $\xi_{k}(P)=\xi
_{k+q}\left(  0\right)  \ $where the indices are taken modulo $K.$
\end{prop}

Figure \ref{fig:1} and Table~\ref{table:ener} show some case examples of these
steady states. We now provide the layout of the results that follow. Firstly,
we give the full characterization of stability of these filament rings, both
energetic and dynamical. This is done in Section \ref{sec:stabring}. We then
consider more general helical states where each z-cross-section is a steady
state corresponding to (\ref{vortex-ss}). We refer to such a state as a
helical filament lattice. For a general vortex lattice that is not on a ring,
we consider only periodic boundary conditions (so that $q$ is a multiple of
$K$ in (\ref{1244}, \ref{BC}, \ref{omega})). In Section \ref{sec:lattice} we
show that such a filament lattice is dynamically stable, provided that the
underlying vortex lattice is stable. However it can become energetically
unstable for sufficiently small $\Omega.$ Some direct numerical simulations of
the proposed equilibria are given in~\ref{sec:comp}, exploring our findings in
the full 3D model, before raising some questions for future study in
Section~\ref{sec:discuss}.

\label{sec:ener}\begin{table}[tb]
\begin{center}%
\begin{tabular}
[c]{l|ccccccc}
& $q=0$ & $q=1$ & $q=2$ & $q=3$ & $q=4$ & $q=5$ & $q=6$\\\hline
&  &  &  &  &  &  & \\
$K=2$ & \torg{20} & \torg{21} & \torg{22} & \torg{23} & \torg{24} &
\torg{25} & \torg{26}\\
$s_{\max,e}$ & $\infty$ & 4 & 0.6667 & 0.25 & 0.1333 & 0.08333 & 0.05714\\
$(m,M)$ &  & (1,0) & (0,1) & (0,1) & (0,1) & (0,1) & (0,1)\\\hline
&  &  &  &  &  &  & \\
$K=3$ & \torg{30} & \torg{31} & \torg{32} & \torg{33} & \torg{34} &
\torg{35} & \torg{36}\\
$s_{\max,e}$ & $\infty$ & 18 & 4.5 & 1.333 & 0.6545 & 0.3956 & 0.2667\\
$(m,M)$ &  & (1,0) & (1,0) & (0,1) & (0,1) & (0,1) & (0,1)\\\hline
&  &  &  &  &  &  & \\
$K=4$ & \torg{40} & \torg{41} & \torg{42} & \torg{43} & \torg{44} &
\torg{45} & \torg{46}\\
$s_{\max,e}$ & $\infty$ & 48 & 12 & 4.8 & 2 & 1.143 & 0.75\\
$(m,M)$ &  & (1,0) & (1,-1) & (0,1) & (0,1) & (0,1) & (0,1)\\\hline
$K=5$ &  &  &  &  &  &  & \\
$s_{\max,e}$ & $\infty$ & 100 & 25 & 11.11 & 5.128 & 2.667 & 1.681\\
$(m,M)$ &  & (1,0) & (1,0) & (1,0) & (0,1) & (0,1) & (0,1)\\\hline
$K=6$ &  &  &  &  &  &  & \\
$s_{\max,e}$ & $\infty$ & 180 & 40.63 & 16.28 & 10.16 & 5.625 & 3.333\\
$(m,M)$ &  & (1,0) & (2,-1) & (3,-2) & (2,-2) & (0,1) & (0,1)\\\hline
$K=7$ &  &  &  &  &  &  & \\
$s_{\max,e}$ & $\infty$ & 294 & 39.2 & 18.38 & 9.8 & 5.939 & 4.356\\
$(m,M)$ &  & (1,0) & (3,-1) & (3,-1) & (3,-2) & (3,-2) & (3,-3)
\end{tabular}
\end{center}
\caption{Energetic stability classification with $K\leq7,q\leq6$. The ring is
stable when $0<s\leq s_{\max,e}$ and is unstable otherwise.}%
\label{table:ener}%
\end{table}

\section{Stability of filament ring states\label{sec:stabring}}

To analyze the ring stability, we deploy the complex variables-based technique
of \cite{kolokolnikov2014tale, Kolokolnikov:2010} in order to examine the
circular Fourier modes of a ring. We start with a general perturbation of the
$\left(  K,q\right)  $ ring state as follows:%

\begin{equation}
\xi_{k}(t,z)=e^{i2\pi k/K}\left(  e^{i\omega z}r+\phi_{k}(t,z)\right)  ,\text{
\ \ }\phi_{k}\ll1.\label{pert}%
\end{equation}
This yields the following linear system for $\phi_{k}:$
\[
\left(  -i\gamma_{1}+\gamma_{2}\right)  \phi_{k}^{\prime}=\left(
D\partial_{zz}-\Omega\right)  \phi_{k}+e^{i\omega z}\sum_{j\neq k}\frac
{1}{r^{2}}\frac{1}{4\sin^{2}\left(  \pi\left(  j-k\right)  /K\right)  }\left(
\exp\left(  2\pi i\left(  j-k\right)  /K\right)  \bar{\phi}_{k}-\bar{\phi}%
_{j}\right)  .
\]
Next, we decompose the perturbation into Fourier modes using the following
self-consistent anzatz:
\begin{equation}
\phi_{k}=e^{i\left(  \alpha+\omega\right)  z}e^{2\pi imk/K}\phi_{+}%
(t)+e^{i\left(  -\alpha+\omega\right)  z}e^{-2\pi imk/K}\bar{\phi}_{-}(t).
\end{equation}
Collecting the like terms in $e^{i\left(  \alpha+\omega\right)  z}e^{2\pi
imk/K}$ and $e^{i\left(  -\alpha+\omega\right)  z}e^{-2\pi imk/K}$ yields a
2x2 system:%
\[%
\begin{array}
[c]{c}%
\left(  -i\gamma_{1}+\gamma_{2}\right)  \phi_{+}^{\prime}=\left(  -D\left(
\alpha+\omega\right)  ^{2}-\Omega\right)  \phi_{+}+\sigma_{+}\phi_{-}\\
\left(  +i\gamma_{1}+\gamma_{2}\right)  \phi_{-}^{\prime}=\left(  -D\left(
\omega-\alpha\right)  ^{2}-\Omega\right)  \phi_{-}+\sigma_{+}\phi_{+}%
\end{array}
\]
where%
\begin{equation}
\sigma_{\pm}=\sum_{j=1}^{K-1}\frac{1}{r^{2}}\frac{1}{4\sin^{2}\left(  \pi
j/K\right)  }\left(  \exp\left(  2\pi ij/K\right)  -\exp\left(  \pm2\pi
ijm/K\right)  \right)  .
\end{equation}
Using identities from \cite{kolokolnikov2014tale} (see (3.8)\ there), we
obtain
\begin{equation}
\sigma_{+}=\sigma_{-}=\sigma=\frac{1}{2r^{2}}\left(  m-1\right)  \left(
K-m-1\right)  .\label{sig}%
\end{equation}
Upon substituting%
\[
\phi_{\pm}(t)=e^{\lambda t}\varepsilon_{\pm}%
\]
and using (\ref{7:52d})\ to rewite $D\left(  \alpha\pm\omega\right)
^{2}+\Omega=\frac{K-1}{2r^{2}}+D\alpha^{2}\pm2D\omega\alpha$ we obtain a 2x2
eigenvalue problem\bes\label{eigall}%
\begin{equation}
\left(
\begin{array}
[c]{cc}%
\gamma_{2}-i\gamma_{1} & 0\\
0 & \gamma_{2}+i\gamma_{1}%
\end{array}
\right)  \left(
\begin{array}
[c]{c}%
\varepsilon_{+}\\
\varepsilon_{-}%
\end{array}
\right)  \lambda=\left(
\begin{array}
[c]{cc}%
-\delta_{+} & \sigma\\
\sigma & -\delta_{-}%
\end{array}
\right)  \left(
\begin{array}
[c]{c}%
\varepsilon_{+}\\
\varepsilon_{-}%
\end{array}
\right)  ,\label{eig}%
\end{equation}%
\begin{equation}
\ \delta_{\pm}=\frac{K-1}{2r^{2}}+D\alpha^{2}\pm2D\omega\alpha.\label{1259}%
\end{equation}
\ees

Next, recall that $\xi_{k}(P)=\xi_{k+q}(0).$ From (\ref{pert}), this implies
that
\[
\phi_{k}(P,t)=e^{i2\pi q/K}\phi_{k+q}(0,t)
\]
so that%

\begin{equation}
e^{i\left(  \alpha+\omega\right)  P}e_{+}^{2\pi imk/K}\phi_{+}(t)+e^{i\left(
-\alpha+\omega\right)  P}e^{-2\pi imk/K}\bar{\phi}_{-}(t)=e^{i2\pi q/K}\left(
e_{+}^{2\pi im(k+q)/K}\phi_{+}(t)+e_{-}^{-2\pi im\left(  k+q\right)  /K}%
\bar{\phi}_{-}(t)\right)  . \label{337}%
\end{equation}

Upon substituting $\omega P=2\pi q/K,$ we find that (\ref{337})\ is satisfied
for all $k$ if and only
\begin{equation}
\alpha/\omega=m+MK/q,\ \ \ \ \text{where }M\in%
\mathbb{Z}
.
\end{equation}
We summarize these computations as follows.

\begin{prop}
\label{prop:general}\textbf{[General Stability Formulation]. }Consider the
$(K,q)$ ring state as given by Proposition \ref{prop:ss}. Its stability is
determined by a sequence of $2\times2$ eigenvalue problems (\ref{eigall})
where%
\[
\delta_{\pm}=\frac{K-1}{2r^{2}}+D\left(  \frac{2\pi}{P}\right)  ^{2}\left(
\frac{mq}{K}+M\right)  \left(  \frac{mq\pm2q}{K}+M\right)  ,\ \ \ \sigma
=\frac{1}{2r^{2}}\left(  m-1\right)  \left(  K-m-1\right)  .
\]
Here, $m$ is the azimuthal mode between the $K$ filaments, whereas $M$ is the
Fourier mode along each of the filaments, with the peturbation having the form%
\begin{equation}
\phi_{k}(t,z)=\exp\left(  i\frac{2\pi z}{P}\left(  \frac{mq+q}{K}+M\right)
+2\pi imk/K\right)  \phi_{+}(t)+\exp\left(  i\frac{2\pi z}{P}\left(
\frac{-mq+q}{K}-M\right)  -2\pi imk/K\right)  \bar{\phi}_{-}(t).
\end{equation}

\end{prop}

We now apply this general formula to two specific cases, namely $\gamma_{2}=0$
to which we refer to as \emph{dynamical stability} (for the Hamiltonian case)
or $\gamma_{1}=0$ which we refer to as \emph{energetic stability} (for the
gradient system scenario).

\subsection{Dynamical stability\label{sec:dyn}}

\begin{figure}[tb]
\begin{center}%
\[
\includegraphics[width=1\textwidth]{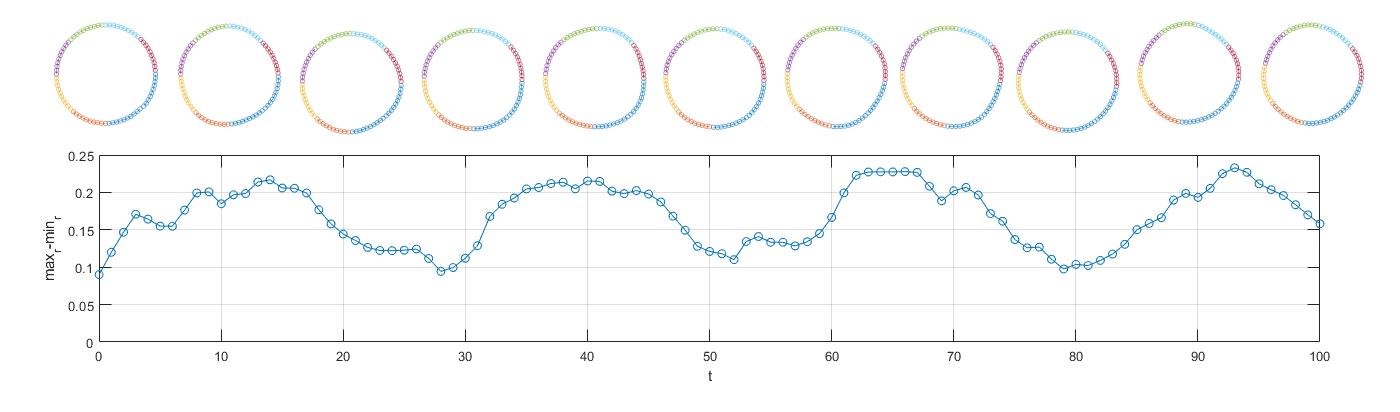}
\]%
\[
\includegraphics[width=1\textwidth]{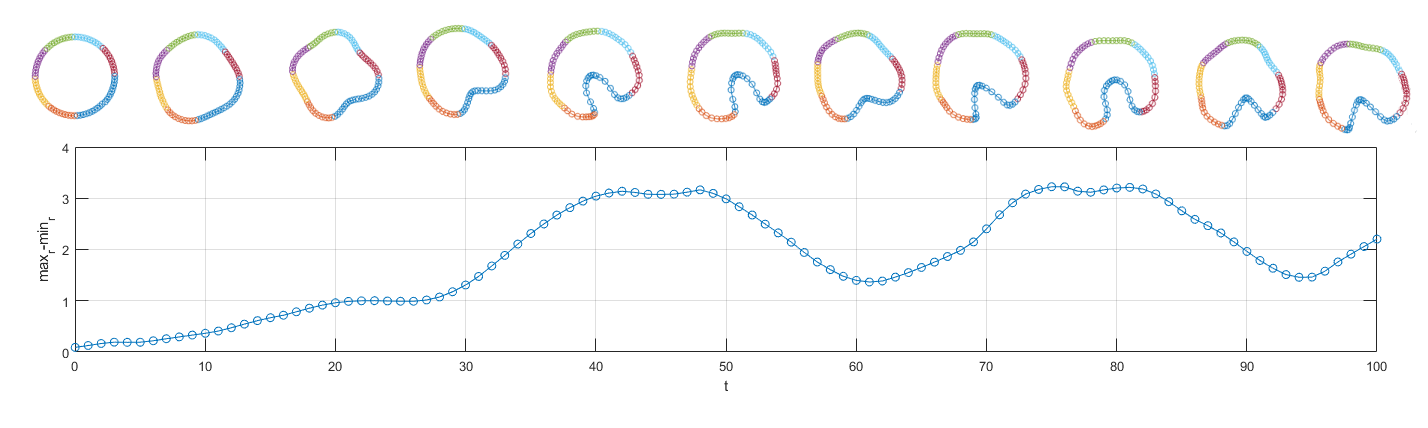}
\]
\end{center}
\caption{Simulation of system (\ref{reduced}) with $K=8,q=1,\ \gamma_{1}=1,$
$\gamma_{2}=0,\ D=1/2$. Top:\ $r=3.0$ (stable regime). Bottom:\ $r=2.5$
(unstable regime). The theoretical stability threshold is $r=\sqrt{8}=2.828.$
The vertical axis shows $\max\left\vert \xi\right\vert -\min\left\vert
\xi\right\vert ,$ versus time $t.$ Snapshots show the actual solution viewed
from the top with $t=0,10,20\ldots100$, with different colours corresponding
to different filaments. \ Initial conditions consist of the the steady state,
slightly perturbed at $t=0.$ Note that the destabilizing mode $m=4$ is clearly
visible (bottom figure)}%
\label{fig2}%
\end{figure}

To study dynamical stability (i.e. stability with respect to (\ref{dyn})), we
set $\gamma_{1}=1,\gamma_{2}=0$ in Proposition \ref{prop:general}. Then
$\lambda$ satisfies%
\begin{equation}
\left(
\begin{array}
[c]{c}%
\varepsilon_{+}\\
\varepsilon_{-}%
\end{array}
\right)  \lambda=\left(
\begin{array}
[c]{cc}%
\delta_{+}i & \sigma i\\
-\sigma i & -\delta_{-}i
\end{array}
\right)  \left(
\begin{array}
[c]{c}%
\varepsilon_{+}\\
\varepsilon_{-}%
\end{array}
\right)  .
\end{equation}
so that
\[
\lambda^{2}-i\left(  \delta_{-}-\delta_{+}\right)  \lambda+\left(  \delta
_{+}\delta_{-}-\sigma^{2}\right)  =0,
\]
having two solutions,
\[
\lambda_{\pm}=\frac{i}{2}\left(  \delta_{-}-\delta_{+}\right)  \pm\frac{1}%
{2}\sqrt{4\sigma^{2}-\left(  \delta+\delta_{+}\right)  ^{2}}.
\]
It follows that the filament ring is dynamically stable if and only if%
\begin{equation}
\left\vert \sigma\right\vert \leq\frac{\delta_{+}+\delta_{-}}{2}. \label{1300}%
\end{equation}
In this case, the eigenvalues $\lambda_{\pm}$ are purely imaginary. Otherwise,
the steady state has saddle structure (with $\operatorname{Re}(\lambda_{+})>0$
and $\operatorname{Re}(\lambda_{-})<0)$. Recalling (\ref{sig}, \ref{1259}),
the stability condition (\ref{1300}) is equivalent to:\
\begin{equation}
K-1+2r^{2}D\left(  \frac{2\pi}{P}\right)  ^{2}\left(  \frac{mq}{K}+M\right)
^{2}\geq\left(  m-1\right)  \left(  K-m-1\right)  \text{, for all }%
m\in\left\{  1\ldots K-1\right\}  ,M\in%
\mathbb{Z}
. \label{1301}%
\end{equation}
We therefore define a dimensionless parameter%
\begin{equation}
s:=\left(  \frac{2\pi}{P}\right)  ^{2}2Dr^{2}. \label{s}%
\end{equation}
so that (\ref{1301}) can be written as $s\geq\frac{K(m-2)-m^{2}+2}{\left(
\frac{mq}{K}+M\right)  ^{2}}.$ We summarize as follows.

\begin{theorem}
\textbf{[Dynamical stability]. }Let
\begin{equation}
s_{\min,d}:=\max_{\substack{m\in\left\{  1\ldots K-1\right\}  ,\\M\in%
\mathbb{Z}
}}\frac{K(m-2)-m^{2}+2}{\left(  \frac{mq}{K}+M\right)  ^{2}}%
\end{equation}
\bigskip Then a $\left(  K,q\right)  $ ring is \textbf{dynamically stable
}privided that $s>s_{\min,d}.$ where $s$ is the dimensionless parameter given
by (\ref{s}).
\end{theorem}

There are several important subcases.

\begin{itemize}
\item \textbf{Case 1:\ }$K\leq7,$\textbf{ any }$q.$ In this case,
$K(m-2)-m^{2}+2\leq0$ for all $m\in\left\{  0,\ldots K-1\right\}  $ so that
$s_{\min,d}\leq0$, and it follows that a ring of $K\leq7$ filaments is\textbf{
dynamically stable} for all $r.$

\item \textbf{Case 2:\ }$K\geq8$\textbf{ and }$\gcd(q$\textbf{,}$K)\neq1.$
Then there are integers $m,M$ with $m\in(1,K-1)$ such that $\frac{mq}{K}+M=0,$
while $K(m-2)-m^{2}+2>0.$ This implies $s_{\min,d}=\infty;$ hence the ring is
\textbf{dynamically unstable} for any $r.$

\item \textbf{Case 3:\ }$K\geq8,$ \textbf{and }$q=1.$ The most unstable mode
turns out to be $\left(  m,M\right)  =\left(  4,0\right)  ,$ with $s_{\min
,d}=\left(  \frac{K-7}{8}\right)  K^{2}.$

Note also that $s_{\min,d}$ does not change when adding any multiple of $K$ to
$q.$ Therefore we may assume without loss of generality that $q\in\left\{
0\ldots K-1\right\}  .$ Table \ref{table:0} lists $s_{\min,d}$ the and
associated destabilizing mode $\left(  m,M\right)  $ for several values of $K$
and $q.$\begin{table}[tb]
\begin{center}%
\[
K=8:\ \ \ \ \
\begin{array}
[c]{c|ccccccccc}%
q & 0 & 1 & 2 & 3 & 4 & 5 & 6 & 7 & 8\\\hline
s_{\min,d} & \infty & 8 & \infty & 64 & \infty & 64 & \infty & 8 &
\infty\\\hline
m &  & 4 & 4 & 3 & 4 & 3 & 4 & 4 & \\\hline
M &  & 0 & -1 & -1 & -2 & -2 & -3 & -4 &
\end{array}
\]%
\[
K=9:\ \ \ \ \
\begin{array}
[c]{c|cccccccccc}%
q & 0 & 1 & 2 & 3 & 4 & 5 & 6 & 7 & 8 & 9\\\hline
s_{\min,d} & \infty & 20\frac{1}{4} & 324 & \infty & 81 & 81 & \infty & 324 &
20\frac{1}{4} & \infty\\\hline
m &  & 4 & 4 & 3 & 4 & 4 & 3 & 4 & 4 & \\\hline
M &  & 0 & -1 & -1 & -2 & -2 & -2 & -3 & -4 &
\end{array}
\]%
\[
K=10:\ \ \ \ \
\begin{array}
[c]{c|ccccccccccc}%
q & 0 & 1 & 2 & 3 & 4 & 5 & 6 & 7 & 8 & 9 & 10\\\hline
s_{\min,d} & \infty & 37\frac{1}{2} & \infty & 300 & \infty & \infty & \infty
& 300 & \infty & 37\frac{1}{2} & \infty\\\hline
m &  & 4 & 5 & 3 & 5 & 4 & 5 & 7 & 5 & 4 & \\\hline
M &  & 0 & -1 & -1 & -2 & -2 & -3 & -5 & -4 & -4 &
\end{array}
\]%
\[
K=11:\ \ \ \ \
\begin{array}
[c]{c|cccccccccccc}%
q & 0 & 1 & 2 & 3 & 4 & 5 & 6 & 7 & 8 & 9 & 10 & 11\\\hline
s_{\min,d} & \infty & 60\frac{1}{2} & 1210 & 968 & 484 & 242 & 242 & 484 &
968 & 1210 & 60\frac{1}{2} & \infty\\\hline
m &  & 4 & 6 & 4 & 3 & 7 & 4 & 3 & 4 & 5 & 7 & \\\hline
M &  & 0 & -1 & -1 & -1 & -3 & -2 & -2 & -3 & -4 & -6 &
\end{array}
\]
\end{center}
\caption{Dynamical stability for filament rings with $K=8,9,10,11$. Stability
range is $s\geq s_{\min,d}$ where $s=\left(  \frac{2\pi}{P}\right)
^{2}2Dr^{2}.$}%
\label{table:0}%
\end{table}
\end{itemize}

\textbf{Example. }Take $K=8$, $q=1.$ Then $s_{\min,d}=8.$ Figure \ref{fig2}
shows the numerical simulations of (\ref{dyn}), with $s$ to either side of the
stability boundary, in full agreement with the predicted stability.

\subsection{Energetic stability}

Energetic stability corresponds to the study of the eigenvalue problem
(\ref{eig}) with $\gamma_{2}=1,\gamma_{1}=0.$ In this case, the eigenvalues
are purely real since they are the eigenvalues of the symmetric matrix
$\left(
\begin{array}
[c]{cc}%
-\delta_{+} & \sigma\\
\sigma & -\delta_{-}%
\end{array}
\right)  .$ Moreover, the trace $-\delta_{+}-\delta_{-}$ is negative, see
(\ref{1259}). It follows that the filament ring is dynamically stable if and
only if determinant is positive, or
\begin{equation}
\sigma^{2}\leq\delta_{+}\delta_{-}. \label{157}%
\end{equation}
Contrast this with the condition for dynamical stability (\ref{1300}%
):\ $\left\vert \sigma\right\vert \leq\frac{\delta_{+}+\delta_{-}}{2}.$ By the
elementary inequality, $\sqrt{\delta_{+}\delta_{-}}\leq\frac{\delta_{+}%
+\delta_{-}}{2}$ for any $\delta_{\pm}>0,$ with equality if and only if
$\delta_{+}=\delta_{-}$, it immediately follows that \textbf{energetic
stability implies dynamical stability. } The former is associated with the
geometric mean of $\delta_{+}$ and $\delta_{-}$, while the latter with the
arithmetic mean thereof. Of course the converse is false since in general,
$\delta_{+}\neq\delta_{-}.$ (the exception is when either $q=0$ or
$m+MK/q=0).$

Written in dimensionless variable $s=\left(  \frac{2\pi}{P}\right)
^{2}2Dr^{2},$ the stability criterion (\ref{157})\ is equivalent to $\mu\geq0$
where%
\begin{equation}
\mu:=\left[  1+\frac{s}{K-1}\left(  \frac{mq}{K}+M\right)  \left(
\frac{mq+2q}{K}+M\right)  \right]  \left[  1+\frac{s}{K-1}\left(  \frac{mq}%
{K}+M\right)  \left(  \frac{mq-2q}{K}+M\right)  \right]  -\left(  m-1\right)
^{2}\left(  1-\frac{m}{K-1}\right)  ^{2}, \label{mu}%
\end{equation}

with the stability boundary corresponding to $\mu=0.$ We summarize as follows.

\begin{prop}
\textbf{[Energetic stability]. }The $\left(  K,q\right)  $ ring is
\textbf{energetically stable }privided that $\mu\geq0$ for all $m\in\left\{
0,1,\ldots K-1\right\}  $ and all $M\in%
\mathbb{Z}
,$ with $\mu$ given by (\ref{mu}). It is unstable otherwise. A ring is
dynamically stable if it is energetically stable (but the converse is not true
in general).
\end{prop}

For further insight, first consider the case $s\rightarrow0.$ One can think of
this as the limit where the Laplacian term in Eq.~(\ref{dyn}) is absent, as is
the case, e.g., for point vortices (rather than filaments). Then,
\begin{equation}
\mu\sim1-\left(  \left(  m-1\right)  \left(  1-\frac{m}{K-1}\right)  \right)
^{2},\ \ \ s=0
\end{equation}
and the stability is independent of $M$ or $q$.
In this case, as is well-known for point vortices~\cite{havelock}, the ring is
stable if $K\leq7$ and is unstable otherwise. So the case $K\leq7$ and
$K\geq7$ must be analyzed separately.

Next, consider the mode $\left(  m,M\right)  =\left(  1,0\right)  $. Then
$\mu$ simplifies to%
\begin{equation}
\mu=\left(  1-s\frac{q^{2}}{\left(  K-1\right)  K^{2}}\right)  \left(
1+3s\frac{q^{2}}{\left(  K-1\right)  K^{2}}\right)  ,\ \ \left(  m,M\right)
=\left(  1,0\right)  .
\end{equation}
Therefore this mode is ustable when $s>s_{\left(  1,0\right)  }$ where
\begin{equation}
s_{\left(  1,0\right)  }:=\frac{\left(  K-1\right)  K^{2}}{q^{2}}.
\end{equation}
In fact, from (\ref{7:52d}, \ref{s})\ note that the rotation rate $\Omega$ can
be written
as $\Omega=\frac{2\pi}{P}D\left(  \frac{K-1}{s}-\frac{q^{2}}{K^{2}}\right)  .$
Thus the threshold $s=\left(  K-1\right)  K^{2}/q^{2}$ corresponds
\emph{precisely }to the zero-rotation rate $\Omega=0$.

Next, take $\left(  m,M\right)  =\left(  0,1\right)  $ in which case we obtain%
\begin{equation}
\mu=\frac{s}{\left(  K-1\right)  ^{2}K^{2}}\left[  \left(  K^{2}%
-4q^{2}\right)  s+2\left(  K-1\right)  K^{2}\right]  ,\ \ \left(  m,M\right)
=\left(  0,1\right)
\end{equation}
When $q<K/2,$ this mode is always stable. On the other hand, this mode is
unstable if $q>K/2$ and $s>s_{(0,1)}$ where%
\begin{equation}
s_{\left(  0,1\right)  }:=\frac{2\left(  K-1\right)  K^{2}}{4q^{2}-K^{2}}%
\end{equation}
A bit of algebra shows that $s_{\left(  0,1\right)  }<s_{\left(  1,0\right)
}$ whenever $K/2<q<K/\sqrt{2}.$

Table \ref{table:1} list the value of $s_{\max,e}$ and the corresponding mode
$(m,M)$ for $2\leq K\leq7$ and $0\leq q\leq6.$ With some exceptions (such as
$\left(  K,q\right)  =\left(  6,3\right)  $ or $K=7,$ $q\geq2$), the
instability threshold corresponds to either $s_{\left(  0,1\right)  }$ or
$s_{\left(  1,0\right)  }$.

Finally, consider the case $K>7.$ Then additional algebra shows that $\left(
K,q\right)  $ with $q\neq1$ is unstable for \emph{all} $s\geq0.$ On the other
hand, when $q=1,$ there exists a stability band $s_{\min,e}\leq s\leq
s_{\max,e}$ where it is stable, while it is unstable outside this range. The
upper bound corresponds to the mode $\left(  m,M\right)  =\left(  1,0\right)
$ given by $s_{\max,e}=s_{(1,0)}.$ On the other hand, additional computations
reveal that the lower bound corresponds to the mode $\left(  m,M\right)
=\left(  4,0\right)  .$ Upon substituting $\left(  m,M\right)  =\left(
4,0\right)  $ and setting $\mu=0,$ we find that $s_{\max,e}$ is the positive
root of%
\begin{equation}
s^{2}+\frac{1}{6}K^{2}\left(  K-1\right)  s-\frac{1}{24}K^{4}\left(
K-4\right)  \left(  K-7\right)  =0 \label{m4}%
\end{equation}

For large $K,$ this asymptotes to $s_{e}\sim K^{3}\frac{\sqrt{7}-1}{12}.$ We
now summarize the above discussion as follows.

\begin{prop}
The $\left(  K,q\right)  $ ring with $q>0$ is \textbf{energetically unstable
}with respect to mode $\left(  m,M\right)  =\left(  1,0\right)  $ when
$s>s_{\left(  1,0\right)  }$, or equivalently, when $\Omega>0.$ The threshold
$s=s_{(1,0)}$ corresponds to rotation rate $\Omega=0$ so that such ring is
energetically unstable when $\Omega<0.$

Suppose that $K\leq7.$ Then there exists $s_{\max,e}$ such that a ring is
stable if and only if $0<s<s_{\max,e}$. When $K/2<q<K/\sqrt{2},$ $s_{\max
,e}\leq s_{\left(  0,1\right)  }<s_{(1,0)}.$ Table \ref{table:1} reports
$s_{\max,e}$ for small $q.$

Suppose that $K>7$ and $q=1.$ Then the ring is energetically stable if and
only if $\ s_{\min,e}\leq s\leq s_{\max,e}$ where $s_{\min,e}=s_{(4,0)}$ is
the positive root of (\ref{m4}), corresponding to the mode $\left(
m,M\right)  =\left(  4,0\right)  $, whereas $s_{\max,e}=s_{\left(  1,0\right)
},$ corresponding to the mode $\left(  m,M\right)  =\left(  1,0\right)  $.

Suppose that $K>7$ and $q\neq1.$ Then the ring is energetically unstable.
\end{prop}

\begin{table}[tb]
\begin{center}%
\[
\ \ \ \ \
\begin{array}
[c]{c|cccccccccc}%
K & 8 & 9 & 10 & 11 & 12 & 20 & 50 & 100 & 200 & \gg1\\\hline
s_{\min,d} & 8 & 20.25 & 37.5 & 60.5 & 90 & 650 & 13437 & 116250 & 965000 &
\sim0.125K^{3}\\
s_{\min,e} & 8.234 & 21.164 & 39.564 & 64.237 & 96 & 703.7 & 14677 & 127276 &
1057677 & \sim0.137K^{3}\\
s_{\max,e} & 448 & 648 & 900 & 1210 & 1584 & 7600 & 122500 & 990000 &
7960000 & \sim K^{3}%
\end{array}
\]
\end{center}
\caption{Stability classification for $K>7,$ $q=1$. The ring is energetically
stable iff $s_{\max,e}\leq s\leq s_{\max,e}.$ It is dynamically stable iff
$s_{\min,d}\leq s.$}%
\label{table:1}%
\end{table}

Let us contrast dynamical and energetic stability. When $K\leq7,$ the ring is
dynamically stable for all $q$ and $s$. On the other hand, it is energetically
stable only within the range $0<s<s_{\max,e}$. \ as given in Table
\ref{table:1}. When $K>7$ and $q\neq1,$ the ring is always energetically
unstable. However it can be dynamically stable for $s>s_{\min,d}$ as long as
$\gcd(K,q)=1;$ see Table \ref{table:0}. Finally when $K>7$ and $q=1,$ the ring
is energetically stable only in the range $s_{\min,e}\leq s\leq s_{\max,e}$
whereas it is dynamically stable in the range $s_{\min,d}\leq s.$ Table
\ref{table:1} gives a comparison between $s_{\min,e}$ and $s_{\min,d}.$
Although not equal, these values are close to each other (within 9\% for large
$K$).

\section{Helical filament lattice}

\label{sec:lattice}\begin{figure}[tb]
\begin{center}
$\includegraphics[width=0.98\textwidth]{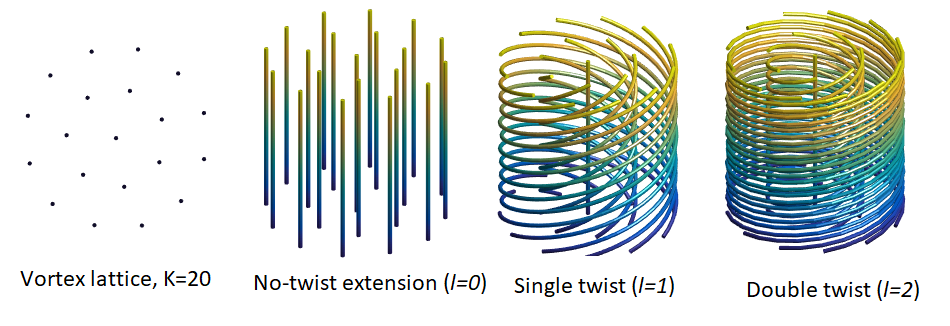}$
\end{center}
\caption{Left:\ stable vortex lattice with 20 point vortices, corresponding to
the stable steady state of (\ref{etap}). The remaining panels show the
extension of this crystal to helical filament crystal with zero, 1 and 2
twists.}%
\label{fig:crystal}%
\end{figure}

Consider \emph{any} relative equilibrium of point vortices $\eta_{k}^{0}$
satisfying (\ref{vortex-ss}). Then $\xi_{k}(z)=e^{zi\omega}\eta_{k}^{0}$
corresponds to a filament equilibrium satisfying (\ref{general}) with a
shifted frequency $\Omega=$ $\Gamma-D\omega^{2}.$ Moreover assume periodic
boundary conditions, so that
\begin{equation}
\omega=\frac{2\pi}{P}l\ \ \text{for an integer }l.
\end{equation}
We refer to such configurations as helical filaments with $l$ twists.

Suppose that the underlying vortex equilibrium is \emph{energetically stable
in the x-y plane. }In other words, $\eta_{k}^{0}$ is a \emph{stable}
equilibrium of the system%
\begin{equation}
\frac{d}{dt}\eta_{k}=-\Gamma\eta_{k}+\sum_{j\neq k}\frac{\eta_{k}-\eta_{j}%
}{\left\vert \eta_{k}-\eta_{j}\right\vert ^{2}}\text{.} \label{etap}%
\end{equation}
A relevant question then is what can be said, in general, about the associated
3D helical extension in terms of stability. By analogy to vortex crystals, we
refer to such filament configurations as helical filament crystals. An example
of a vortex crystal consisting of 20 vortices and its extensions are shown in
Figure \ref{fig:crystal}. We show the following result.

\begin{thm}
\label{prop:app}Let $\eta_{k}^{0}$ be a \textbf{stable} equilibrium of the
system (\ref{etap}). Let $\xi_{k}=\exp\left(  \omega iz\right)  \eta_{k}^{0}$
be the corresponding twisted filament relative equilibrium satisfying%
\begin{equation}
0=D\frac{\partial^{2}}{\partial z^{2}}\xi_{k}-\Omega\xi_{k}+\sum_{j\neq
k}\frac{\xi_{k}-\xi_{j}}{\left\vert \xi_{k}-\xi_{j}\right\vert ^{2}}%
\end{equation}
where $\Gamma=D\omega^{2}+\Omega>0.$ Assume periodic boundary conditions for
$z\in\lbrack0,P],$ so that $\omega=\frac{2\pi}{P}l,$ $l\in%
\mathbb{Z}
.$ \ We have the following:

\begin{itemize}
\item $\xi_{k}$ is energetically stable (i.e. stable with respect to
(\ref{ener}))\ if and only if $\Omega>\Omega_{\min,e}$ where
\begin{equation}
\Omega_{\min,e}:=D\left(  \frac{2\pi}{P}\right)  ^{2}\left(  l^{2}-\frac{1}%
{2}\right)  . \label{Omegamin}%
\end{equation}

\item $\xi_{k}$ is dynamically stable (i.e. stable with respect to
(\ref{dyn}))\ for all $\Omega.$
\end{itemize}
\end{thm}

Note that $\Omega$ could be negative as long as $\Gamma>0.$ The latter
condition is necessary for the steady state to exist, see (\ref{644}). Before
showing \ref{prop:app}, we will need the following lemma.

\begin{lem}
\label{lem:L}Let $\eta_{k}^{0}$ be a \textbf{stable }equilibrium of the system
(\ref{etap}). Let $\phi=\left(  \phi_{1},\ldots\phi_{K}\right)  $ and define
the operator%
\begin{equation}
L\phi=\sum_{j\neq k}\frac{-1}{\left(  \overline{\eta}_{k}^{0}-\overline{\eta
}_{j}^{0}\right)  ^{2}}\left(  \phi_{k}-\phi_{j}\right)  . \label{L}%
\end{equation}
Suppose that $\mu$ is an eigenvalue of $\bar{L}L$, where $\bar{L}$ involves
taking a conjugate of Eq.~(\ref{L}), i.e. $\bar{L}\phi=\sum_{j\neq k}\frac
{-1}{\left(  \eta_{k}^{0}-\eta_{j}^{0}\right)  ^{2}}\left(  \phi_{k}-\phi
_{j}\right)  $. Then $\mu$ satisfies $0\leq\mu\leq\Gamma^{2}.$ Moreover, the
maximum is achieved: there is an eignevalue $\mu=\Gamma^{2}$ of $\bar{L}L$.
\end{lem}

\textbf{Proof of Lemma \ref{lem:L}. }First, note that $L$ is symmetric so that
$\bar{L}L$ is positive definite and hence all of its eigenvalues $\mu\geq0$.
To see that $\max\mu=\Gamma^{2},$ we linearize (\ref{etap})\ around the
equilibrium as $\eta_{k}(t)=\eta_{k}+e^{\kappa t}\phi_{k}$ to obtain the
problem%
\begin{equation}
\left(  \kappa+\Gamma\right)  \phi=L\bar{\phi}. \label{a1}%
\end{equation}
where $L$ is given by (\ref{L}), and $\kappa$ is the eigenvalue of the
linearization of (\ref{etap}). Taking a conjugate, we have%
\[
\left(  \bar{\kappa}+\Gamma\right)  \bar{\phi}=\bar{L}\phi.
\]
Applying $\bar{L}$ to both sides of (\ref{a1})\ yields
\begin{equation}
\left\vert \kappa+\Gamma\right\vert ^{2}\bar{\phi}=\bar{L}L\bar{\phi}%
\end{equation}
Since (\ref{etap}) is the gradient flow of the associated energy
$E=-\Gamma\sum_{k}\frac{\left\vert \eta_{k}\right\vert ^{2}}{2}+\sum
\sum_{j\neq k}\log\left\vert \eta_{k}-\eta_{j}\right\vert $, the relevant
eigenvalues $\kappa$ are all purely real. Therefore we have%
\begin{equation}
\left(  \kappa+\Gamma\right)  ^{2}\bar{\phi}=\bar{L}L\bar{\phi}.
\end{equation}
It follows that $\left(  \kappa+\Gamma\right)  ^{2}=\mu$ for some eigenvalue
$\mu$ of $\bar{L}L.$ Conversely, the matrix $\bar{L}L$ has $K$ eigenvalues
whereas the linearization of problem (\ref{etap}) has $2K$ eigenvalues.
Therefore $\kappa=-\Gamma\pm\sqrt{\mu}$ are both eigenvalues of the
linearization of Eq.~(\ref{etap}) for any given eigenvalue $\mu$ of $\bar{L}%
L$.
Given the stability assumption above for $\eta_{k}^{0}$, it follows that
$\kappa=-\Gamma\pm\sqrt{\mu} \leq0$ or $\mu\leq\Gamma^{2}.$ Finally, the
problem (\ref{etap}) admits a zero eigenvalue $\kappa=0$ corresponding to
rotation invariance, so that $\mu=\Gamma^{2}$ is the maximum eigenvalue of
$\bar{L}L.$ $\blacksquare$

\textbf{Proof of Proposition \ref{prop:app}. }We first prove (a). Linearize
equations (\ref{ener}) as%
\begin{equation}
\xi_{k}(z,t)=\xi_{k}(z)+\phi_{k}(z,t),\ \ \ \phi_{k}\ll1
\end{equation}
to obtain%
\begin{equation}
(\partial_{t}+\Omega-D\partial_{zz})\phi=e^{2i\omega z}L\bar{\phi}%
\end{equation}
where $\phi=\left(  \phi_{1},\ldots\phi_{K}\right)  ^{T}$ and $L$ is the
linear operator (\ref{L}).

Next we use the following anzatz:%
\[
\phi=\phi_{+}(t)e^{i(\omega+\alpha)z}+\bar{\phi}_{-}(t)e^{i(\omega-\alpha)z}.
\]
to obtain%
\begin{equation}
(\partial_{t}+\Omega+D\left(  \omega+\alpha\right)  ^{2})\phi_{+}=L\phi
_{-};\ \ \ \ (\partial_{t}+\Omega+D\left(  \omega-\alpha\right)  ^{2}%
)\bar{\phi}_{-}=L\bar{\phi}_{+};
\end{equation}
Taking complex conjugate of the second equation we get%
\begin{equation}
(\partial_{t}+\Omega+D\left(  \omega-\alpha\right)  ^{2})\phi_{-}=\bar{L}%
\phi_{+}.
\end{equation}
Finally we take
\begin{equation}
\phi_{\pm}=e^{\lambda t}\varepsilon_{\pm}%
\end{equation}
to obtain%
\[
(\lambda+\Omega+D\left(  \omega+\alpha\right)  ^{2})\varepsilon_{+}%
=L\varepsilon_{-},\ \ \ \ (\lambda+\Omega+D\left(  \omega-\alpha\right)
^{2})\varepsilon_{-}=\bar{L}\varepsilon_{+}.
\]
We apply $\bar{L}$ to the first equation o obtain%
\begin{equation}
\left(  \lambda+D\left(  \omega+\alpha\right)  ^{2}+\Omega\right)  \left(
\lambda+D\left(  \omega-\alpha\right)  ^{2}+\Omega\right)  =\mu\label{1139}%
\end{equation}
where $\mu$ is eigenvalue of $\bar{L}L.$ In terms of $\Gamma$ this becomes%
\begin{equation}
\lambda^{2}+2\lambda\left(  \alpha^{2}+\Gamma\right)  +D\alpha^{2}\left(
D\alpha^{2}-4D\omega^{2}+2\Gamma\right)  +\Gamma^{2}-\mu=0.
\end{equation}
Therefore a necessary and sufficient condition for stability is that
$D\alpha^{2}\left(  D\alpha^{2}-4d\omega^{2}+2\Gamma\right)  +\Gamma^{2}%
-\mu\geq0$ for all admissible $\alpha,\mu.$ By Lemma \ref{lem:L}, $0\leq
\mu\leq\Gamma^{2}$ with $\max\mu=\Gamma^{2}.$ So the stability condition
becomes $D\alpha^{2}\left(  D\alpha^{2}-4d\omega^{2}+2\Gamma\right)  \geq
\max(\mu)-\Gamma^{2}=0.$ Upon substituting $\Gamma=D\omega^{2}+\Omega$, this
is equivalent to $D\alpha^{2}-2D\omega^{2}+2\Omega\geq0$, or $\Omega\geq
\max_{\alpha,\alpha\neq0}D\left(  \omega^{2}-\frac{\alpha^{2}}{2}\right)  .$
This is maximized when $M=\pm1,\ \alpha=2\pi/P,$ showing (\ref{Omegamin}).

We now show part (b). Linearizing (\ref{ener})\ in the same way as part\ (a),
we obtain%
\begin{align*}
(-i\partial_{t}+\Omega+D\left(  \omega+\alpha\right)  ^{2})\phi_{+}  &
=L\phi_{-};\\
(i\partial_{t}+\Omega+D\left(  \omega-\alpha\right)  ^{2})\phi_{-}  &
=\bar{L}\phi_{+}.
\end{align*}
and instead of (\ref{1139}) we obtain%
\begin{equation}
\left(  -i\lambda+D\left(  \omega+\alpha\right)  ^{2}+\Omega\right)  \left(
i\lambda+D\left(  \omega-\alpha\right)  ^{2}+\Omega\right)  =\mu
\end{equation}
so that $\lambda$ satisfies%
\begin{equation}
\lambda^{2}+i\lambda4D\omega\alpha+D\alpha^{2}\left(  D\alpha^{2}-4D\omega
^{2}+2\Gamma\right)  +\Gamma^{2}-\mu=0.
\end{equation}
After some algebra we obtain%
\[
\lambda=\left(  -2D\omega\alpha\pm\sqrt{D\alpha^{2}\left(  D\alpha^{2}%
+2\Gamma\right)  +\Gamma^{2}-\mu}\right)  i
\]
By Lemma \ref{lem:L}, $\mu<\Gamma^{2}$ so the expression under the square root
is always positive. This shows that $\lambda$ is purely imaginary for all
parameter values which proves dynamical stability. $\blacksquare$

\begin{figure}[tb]
\begin{center}%
\[
\includegraphics[width=0.98\textwidth]{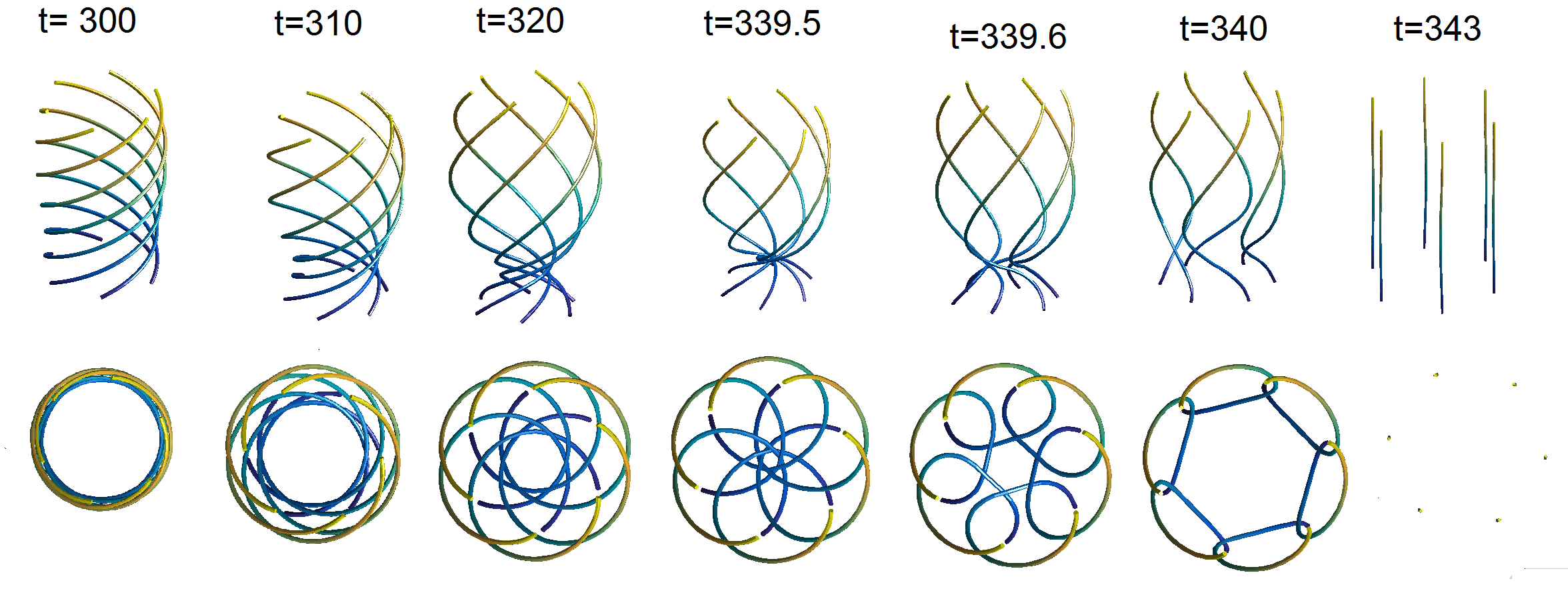}
\]
\[
\includegraphics[width=0.98\textwidth]{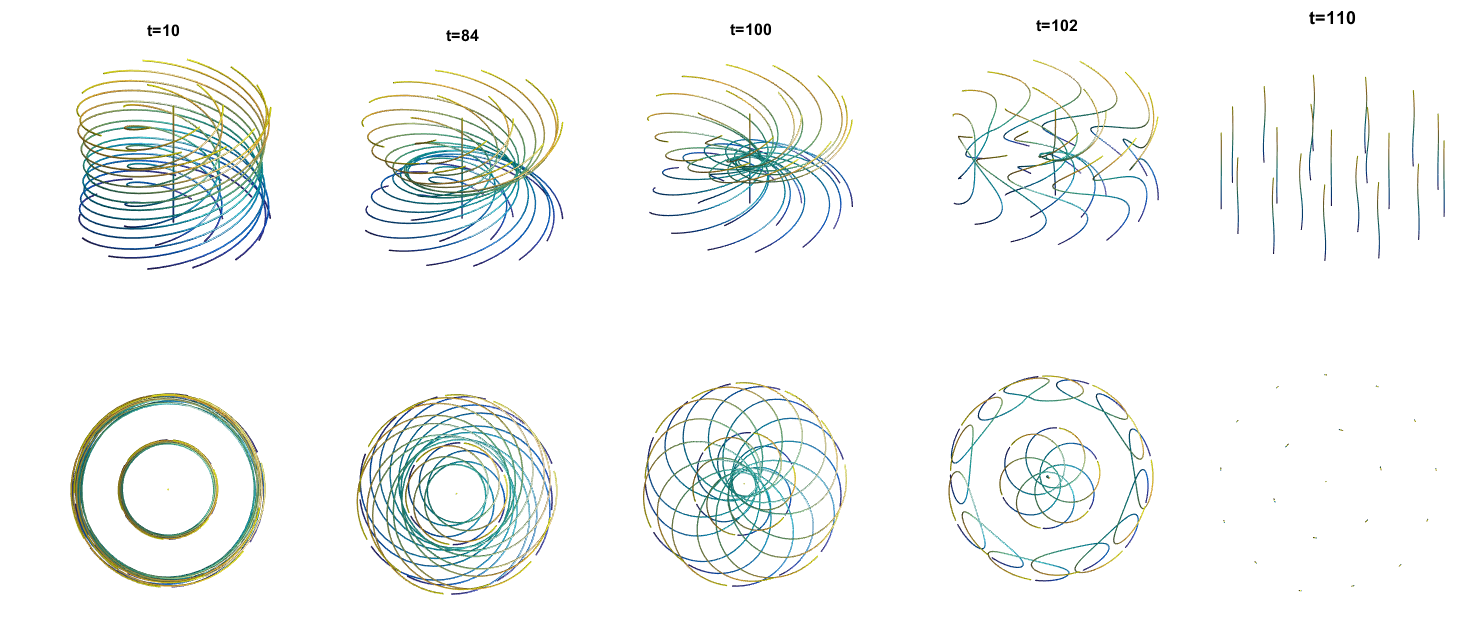}
\]
\end{center}
\caption{Top:\ Simulations of (\ref{dyn})\ starting with a ring $(K,q)=(6,6)$
and with $D=0.5,\ \Omega=0.24.$ Instability is observed, followed by a
finite-time collapse as the filaments cross each-other. Continuing the
simulation after the collapse leads to a stable ring of straight filaments.
Bottom:\ Simulations of (\ref{dyn})\ with $K=20$ filaments and with initial
conditions as in Figure \ref{fig:crystal} with $l=1.$ The remaining parameters
are the same as above.}%
\label{fig:ring6}%
\end{figure}

\textbf{Example: filament ring. } First, we verify that this result agrees
with stability of ring solutions, refer to Figure \ref{fig:ring6}. For
periodic solutions, $l=q/K$ must be an integer. For example, take $K=6,q=6;$
from table II; we read the threshold of $s_{\max,e}=3.33.$ Recalling
(\ref{7:52d}, \ref{s}), \ we have $\Omega=\frac{2\pi}{P}D\left(  \frac{K-1}%
{s}-\frac{q^{2}}{K^{2}}\right)  =D\left(  \frac{2\pi}{P}\right)  ^{2}%
\cdot0.50.$ This coresponds precisely to (\ref{Omegamin})\ with $l=1=q/K.$
Numerical simulations with $D=0.5,P=2\pi,$ $\Omega=0.24$ are shown in Figure
\ref{fig:ring6}~(top). \ For this value, $\Omega<\Omega_{\min,e}=0.25$ and as
expected, an instability is observed. This instability leads to a finite-time
\textquotedblleft collapse\textquotedblright\ around $t\approx339.5$,
corresponding to the crossing of the filaments. Continuing numerical
simulations of (\ref{dyn})\ beyond this collapse leads to another
(stable)\ ring, this time with $l=0.$

\textbf{Example:\ crystal filament lattice. }For more than 7 point vortices,
the energetically preferred state is a \textquotedblleft
lattice\textquotedblright\ such as shown in Figure \ref{fig:crystal}. As in
the preceeding example, taking $P=2\pi,D=1/2,$ we find that $\Omega_{\min
}=0.25$ when $l=1$ (indeed $\Omega_{\min}$ only depends on $l$ and not on the
number of filaments). Simulation of (\ref{dyn})\ with $\Omega=0.26$ and $l=1$
exhibits a stable state. On the other hand, decreasing $\Omega=0.24$ results
in an instability of the $l=1$ lattice, as illustrated in Figure
\ref{fig:ring6}.

We remark that $\Omega_{c}$ is always positive since $\omega\geq2\pi/P$. Note
that the more twisted (bigger $\omega$) the configuration is, the bigger the
rotation $\Omega$ should be to stabilize it. Also for an infinitely long
filament ($P=\infty$), the equation (\ref{Omegamin})\ reduces to $\Omega
_{\min}=0.$

\bigskip

\bigskip

\bigskip

\section{Comparison to full numerical simulations of GP equations}

\label{sec:comp}

In this section we discuss our simulations of the full solution of the
governing GP\ PDE in 3D. Recall that the latter is the natural starting point
for deriving the KMD model, as discussed, e.g., in~\cite{contreras2016nearly}.
The GP equation reads:
\begin{equation}
i\frac{d\psi}{dt}=-\frac{1}{2}\nabla^{2}\psi+V(R)\psi+g|\psi|^{2}\psi.
\label{fullpde}%
\end{equation}
Here $g=4\pi Na/l_{R}$ and $a$ is the s-wave scattering, $l_{R}$ is the axial
oscillator length: $\sqrt{\hbar/m\omega_{R}}$ with $R$ being the axial
coordinate ($R^{2}=x^{2}+y^{2}$), $m$ is the atomic mass, and $\omega_{R}$ is
the axial trap frequency. In our simulations we have rescaled the length by
$l_{R}$, time by $1/\omega_{R}$ and the energy by $\hbar\omega_{R}$ to get the
form in (\ref{fullpde}). Throughout this manuscript, we work in these scaled
dimensions which are tantamount to the dimensions used in the reduced system
of Eq~(\ref{reduced}).
This affords us the opportunity to connect our results to those of the full 3D
field theory. The vortex lines (constituting the filaments of our effective
filament model) are topological defects in the complex order parameter of the
GP theory on which there is a vanishing of the density ($|\psi|$) and around
which a suitable winding of the phase takes place. It is via these features
that we identify and visualize the vortex filaments in the numerical results
described below.

As our choice of trapping for $V(R)$, we simulated both an axial harmonic trap
($\frac{1}{2} R^{2}=\frac{1}{2} (x^{2}+y^{2})$) and a flat bottom trap. We
present the flat bottom trap results to remove the effect of an inhomogeneous
background density profile. This potential has harmonic confinement beyond
$R_{0}$, so $V(R)=1/2(R-R_{0})^{2}\Theta(R-R_{0})$ where $\Theta(x)$ is a
Heaviside function that is 1 when $x\ge0$ and zero otherwise. We have picked
$R_{0}=4$, and we have tested that this does not influence vortex motion when
they are well within this radius (i.e., for $R\le2$).

The chemical potential $\mu$ is chosen to be large (30-42 $\hbar\omega_{r}$),
so this gives a small healing length, $\xi$, as $1/\sqrt{\mu}$. In addition to
$r$ (the ring filament radius) and P (the size of the domain along the $z$
axis), the healing length defines a scale in the simulation which complicates
the comparison with the filament method. We have picked the $r$ to be in the
range of 1-2, so it is much larger than $\xi$ and much smaller than $R_{0}$.
This range of $r$ usually means that the initial separation between vortex
filaments is much larger than $\xi$ and hence the latter scale does not have a
critical role in the dynamics.

The time propagation of Eq.(\ref{fullpde}) takes place with a third-order
operator splitting Fourier spectral method using time steps of $5\times
10^{-4}$. The spatial grid has 128${}^{3}$ points and with a grid spacing of
0.15$l_{R}$ in $x$ and $y$, while $dz$ is varied to get a desired $P$.
We used periodic boundary conditions for all simulations of the vortex twists.
We find the initial condition by first imprinting the phase with the ansatz:
\begin{equation}
\psi(x,y,z)/|\psi|=\Pi_{j}\mbox{tan}^{-1}(x-x_{j},y-y_{j}), \label{phase}%
\end{equation}
where $x_{j}$, $y_{j}$ is the position of the $j$-th vortex core.
Thus, the total phase is simply the sum of all the vortex core phases. After
the phase is defined, we evolve $\psi$ in imaginary time to relax the density.
Once the energy changes by less than $10^{-8}\hbar\omega_{r}$, we consider the
initial condition converged. The phase imprinting locks the vortices in place,
and only the density is changed at this stage, so the configuration (to which
the dynamics locks) can be metastable.

\begin{figure}[ptb]%
\[
\includegraphics[width=0.4\textwidth]{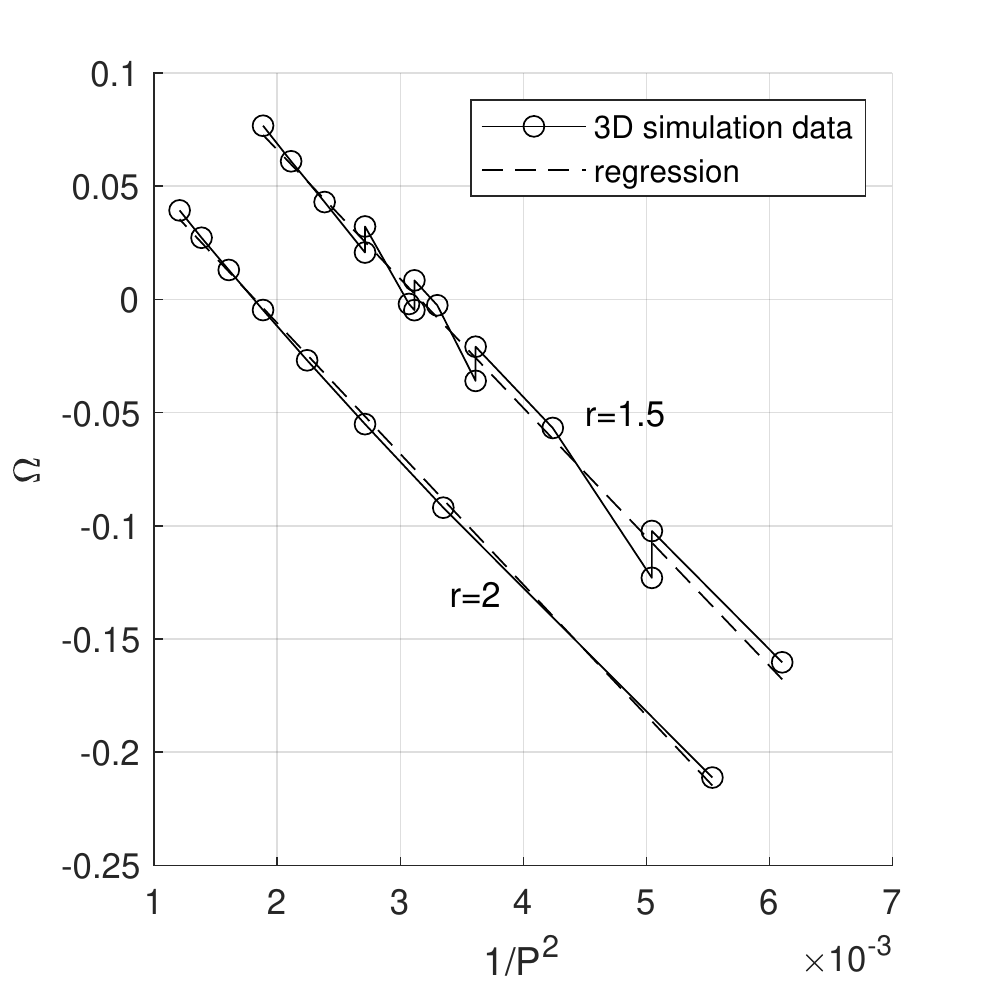}
\]
\caption{Comparison of the full simulation of (\ref{fullpde}) and the theory
(of Eq.~(\ref{7:52d})). Simulations are done for $\left(  K,q\right)  =\left(
2,2\right)  $ ring with fixed $r$ (either $r=1$ or $r=2$), and for several
values of box height $P,$ and the resulting rotation rate $\Omega$ is then
recorded. The theory of Eq.~(\ref{7:52d}) predicts a linear relationship
between $\Omega$ and $P^{-2}$ which is verified in the direct numerical
computations of the 3D GP PDE of Eq.~(\ref{fullpde}).}%
\label{fig:fitomega}%
\end{figure}

To find stationary states in the full simulations we must vary the simulation
box height. An example of this is shown in Fig. \ref{fig:fitomega}, where we
show plot the rotational velocity of the vortex twist as a function of
$1/P^{2}$. Here we can see that $\Omega$ (measured in radians per trap unit of
time) crosses zero at a particular value of $1/P^{2}$. In addition, for r=1.5,
we show the extracted data from simulations with two different chemical
potentials, 30 and 42 $\hbar\omega_{R}$. KMD theory predicts that $\Omega$ is
a linear function of $1/P^{2}$, see (\ref{7:52d}). This is validated with full
GP simulations as seen in Fig. \ref{fig:fitomega}.


We now consider three illustrative examples of for $K=2,3,$ and 4 with $q=K$.
\begin{figure}[tb]
\includegraphics[width=0.3\textwidth]{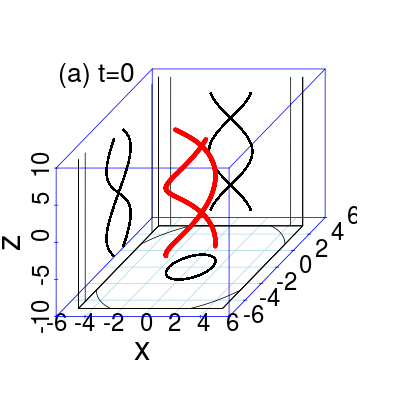}
\includegraphics[width=0.3\textwidth]{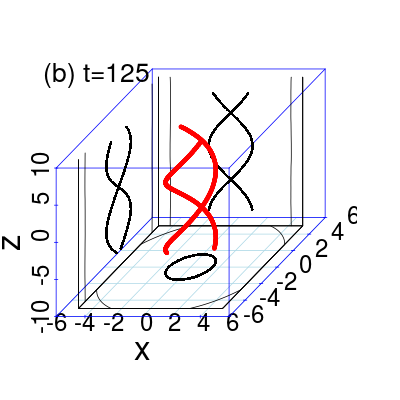}
\includegraphics[width=0.3\textwidth]{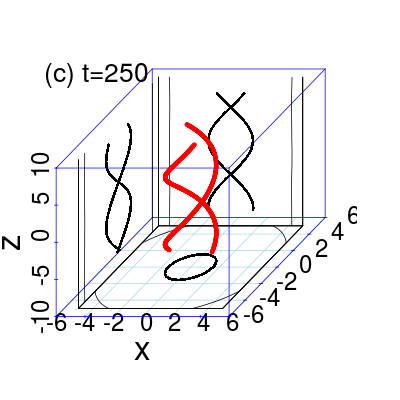}
\includegraphics[width=0.3\textwidth]{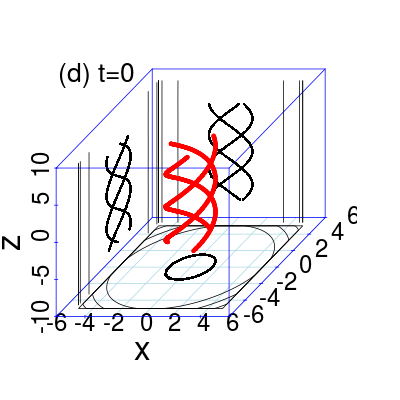}
\includegraphics[width=0.3\textwidth]{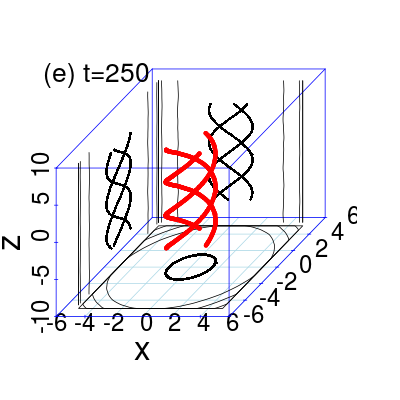}
\includegraphics[width=0.3\textwidth]{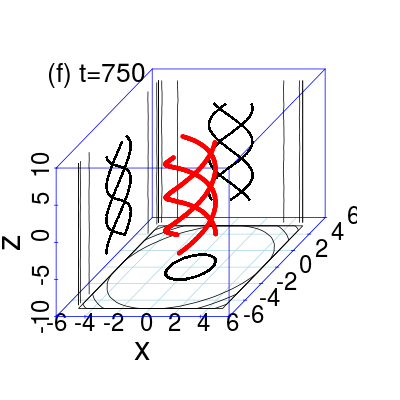}
\caption{For $K=2$ (top) and $K=3$ (bottom), the vortices appear stable and
are very nearly stationary. For $K=2$, the times in trap units are (a) 0, (b)
125, and (c) 250. For $K=3$, the simulations were run for 1.5 million times
steps, three times longer than usual, to test stability. The times shown are
(d) 0, (e) 500 and (f) 750 in trap units. The vortices are shown as red points
and their project is seem in the side as bold black points. The projected
density contours are shown as this black lines on the sides. }%
\label{fig:gpe-23}%
\end{figure}

For $K=2$ and $3$, we were able to find the $P$ that essentially froze the
motion of the vortices. In these figures, the 3D vortex cores are shown as red
lines. The vortex positions are extracted by finding the phase singularity on
the computational grid \cite{FBD2010}. We refine the vortex positions using
methods from Refs. \cite{top-2017,Villois2016}. We also project the vortex
core positions on the sides of the figure. Additionally, we have shown the
projected density of the BEC projected as thin contours on the sides of the
figure. Figure \ref{fig:gpe-23} shows the full simulations for $K=2,3$ with
$q=K.$ Such a filament ring appears to be stable (and practically stationary)
regardless of the initial radius $r.$ This is in agreement with the results in
Section~\ref{sec:dyn}, which show that such a ring is dynamically stable for
all parameters\ (even though it may be energetically unstable, see Section
\ref{sec:ener}). Recall that the Hamiltonian GP model of Eq.~(\ref{fullpde})
is connected at a reduced level with the KMD filament setting of
Eq.~(\ref{reduced}) (rather than with the gradient dynamics of Eq.~(\ref{ener})).

\begin{figure}[tb]
\includegraphics[width=0.3\textwidth]{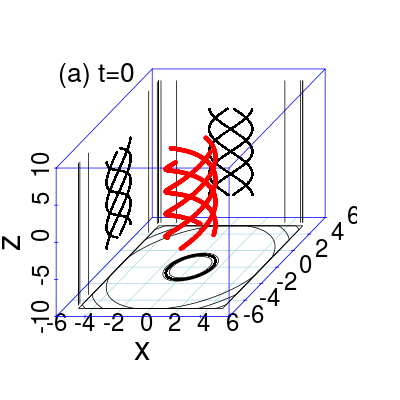}
\includegraphics[width=0.3\textwidth]{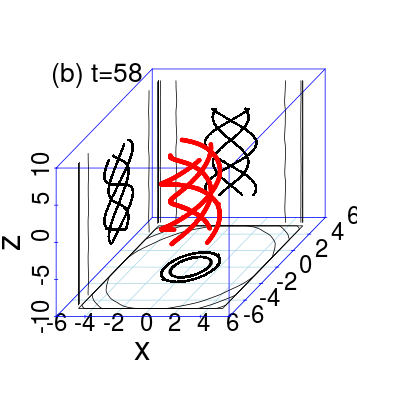}
\includegraphics[width=0.3\textwidth]{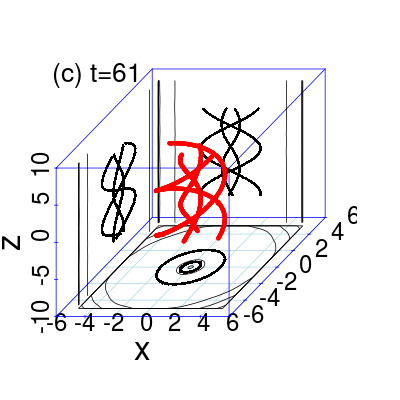}
\includegraphics[width=0.3\textwidth]{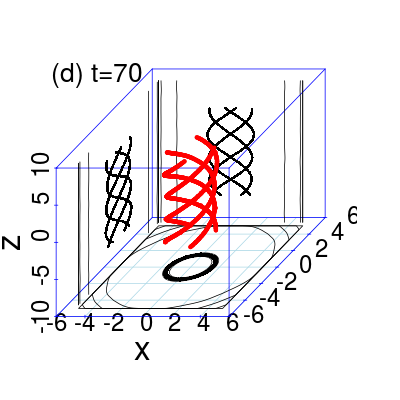}
\includegraphics[width=0.3\textwidth]{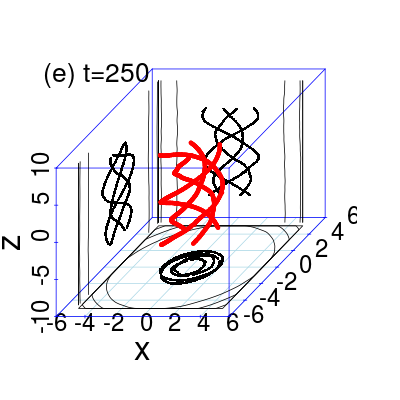}
\includegraphics[width=0.3\textwidth]{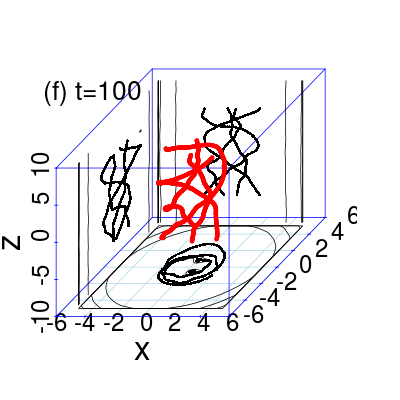}
\par
\caption{For $K=4$, the vortices initialy appear to be stationary, but then a
collective motion breaks out and the vortex configuration scrambles. The times
shown are (a) 0, (b) 58, (b) 61, (d) 70, and (e) 250 in trap units. (e) The
last snapshot is the same configuration with perturbations on the initial
position of the vortices and because of these it quickly evolves out of the
initial configuration. For this simulation the time shown is 100. }%
\label{fig:gpe-42}%
\end{figure}

Figure \ref{fig:gpe-42} shows the full simulations with $K=4,q=4$. The
simulation is initially stationary, but eventually the vortex twist gains a
collective motion that is depicted in the snapshots. This destabilizes the
configuration which is apparently dynamically unstable. Additionally in Figure
\ref{fig:gpe-42}(f), we show the same vortex configuration with slight
perturbations on the initial condition. The end result is that the system
never appears stationary and has a much more disordered appearance. Indeed,
what we expect here is that the correspondence between the KMD filament model
and the complex 3D dynamics of the GP of Eq.~(\ref{fullpde}) may break down as
one goes to a regime involving a large number of filaments rotating at small
distances from one another. Nevertheless, we believe that the above select
examples suggest the analysis of the KMD model as a useful tool for
identifying multi-vortex-filament configurations in the original PDE system of
relevance to a wide range of (e.g., atomic, optical and hydrodynamic) applications.

\section{Discussion}

\label{sec:discuss}

\begin{figure}[ptb]
\caption{Energetically stable configurations of a single vortex filament
wrapped around 20 times (so that $K=20,\xi_{k}(P)=\xi_{k+1\operatorname{mod}%
K}(0)$). The figure shows the view looking down the z-axis (i.e. the
projection onto x-y plane). Here, $\Omega=1$ and $P=2\pi$ with $D$ as
indicated. Each snapshot corresponds to the numerically computed steady state
of (\ref{ener}). For large $D$ $(D>D_{\min}=40.8)$, the steady state is a
ring-like configuration. For small $D$, most x-y cross-sections show a 2D
vortex crystal structure.}%
\label{fig:bif}
\includegraphics[width=0.8\textwidth]{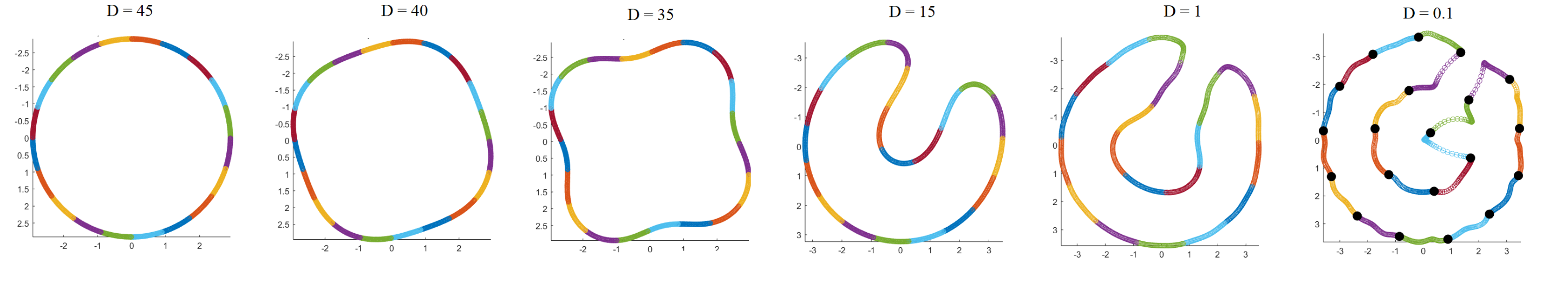}\end{figure}\begin{figure}[tb]
\begin{center}
\includegraphics[width=1\textwidth]{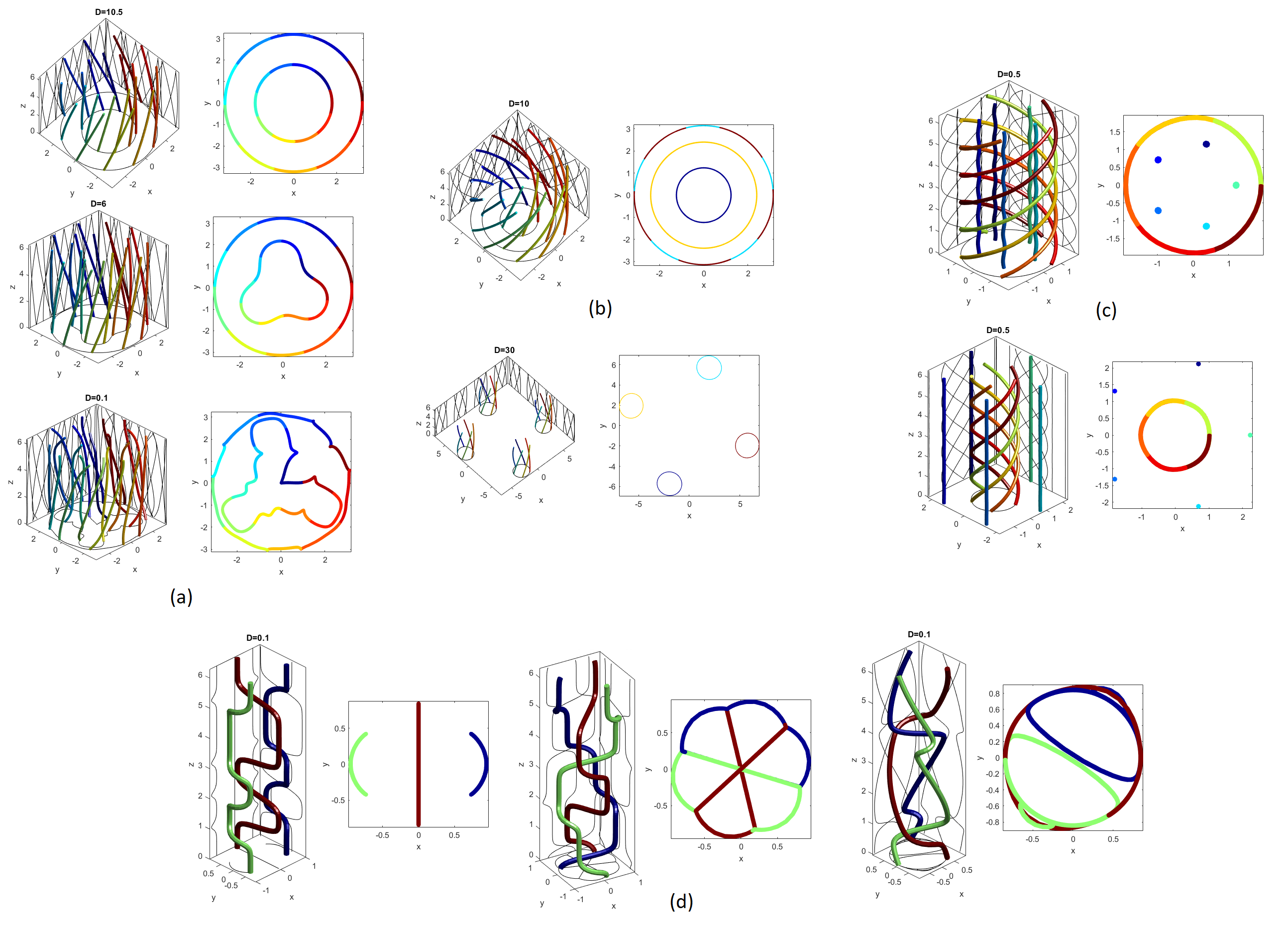}
\end{center}
\caption{A selection of non-helical energetically stable relative equilibria
obtaned by solving (\ref{ener}) numerically.\ (a)\ Inital conditions consist
of a helical ring $\left(  K,q\right)  =(20,2)$ and converge to a double-ring
solution when $D=10.$ A cascade of bifurcations is observed as $D$ is
decreased. (b)\ Taking initial conditions $\left(  K,q\right)  =\left(
20,4\right)  $ with $D=10$ and $D=30$. Note the ``triple-ring'' solution.
(c)\ Mixed helical state (d)\ Three filaments with $D=0.1,$ starting from
different initial conditions. The resulting states exhibit high symmetry and
sharp boundary layers. }%
\label{fig:zoo}%
\end{figure}

Our aim in the present work was to explore the dynamical reduction afforded by
the KMD model in order to propose exact, analytically tractable solutions at
the level of multiple vortex filaments (in particular, filament rings and
lattices). Moreover, this reduced description enabled a systematic
characterization of the stability of such states both at the level of
dynamical stability (of the original Hamiltonian problem), as well as at that
of energetic stability (relevant to the gradient flow of Eq.~(\ref{ener})). We
illustrated that while the more stringent conditions for energetic stability
imply dynamical stability, the converse is false. The relevant conditions of
stability for the vortex filaments depend on the number of filaments (and the
relevant periodicity) with, e.g., $K \leq7$ leading to dynamical stability. On
the other hand, for the helical filament lattices we could establish that the
configuration is dynamically stable provided that the underlying vortex
configuration is energetically stable in the two-dimensional plane. We have
gone on to explore some prototypical ones among these results in a fully
3-dimensional simulation of the original Gross-Pitaevskii model from which the
filament KMD reduction was obtained. We have seen that in some of the simpler
settings involving e.g. 2 or 3 filaments, the results of the 3D simulation are
in line with those of the reduced case. However, for larger numbers of
filaments, we found potential instabilities in the original model that were
not mirrored in the reduction. A systematic exploration of the breakdown of
the model for larger numbers of filaments is a particularly interesting topic
for future study.

Naturally, there are numerous directions that our study paves towards future
work.
For instance, one possibility that we have touched upon and which is
illustrated in Figure \ref{fig:bif} for $K=20,$ $q=1$ is to use $D$ as the
bifurcation parameter (while we fix $\Omega=1$). In that case, we find that
the energetic stability boundaries appear to be {supercritical} and can lead
to bifurcations of novel nearby stable equilibria.
For this particular example, the thresolds $s_{\min,e}=703.7,\ s_{\max
,e}=7600$ (from Table \ref{table:1}) then correspond to $D=40.8$ and
$D=\infty,$ respectively. As $D$ is decreased past $40.8,$ a mode-4
instability destabilizes the ring. This instability appears to be
supercritical:\ while the ring is energetically unstable, a nearby mode-4
pattern appears to be stable, and the system converges to it. As $D$ is
decreased further, subsequent bifurcations are observed. In the limit
$D\rightarrow0,$ the various z- cross-sections decouple and each cross-section
looks like a point vortex lattice; these layers are connected through sharp
transitions. Numerical experiments suggest that bifurcations in Figure
\ref{fig:bif} are {reseversible:\ }as $D$ is increased, the steady state
straightens itself out, eventually resulting in a single ring. Our numerical
experiments suggest the following conjecture worth pursuing in future studies:

\begin{conjecture}
For boundary conditions $\xi_{k}(P)=\xi_{k+1\operatorname{mod}K}(0)$\ in the
limit $D\rightarrow\infty,$ the only energetically stable equilibrium is the
$q=1$ ring.
\end{conjecture}

Furthermore, there are numerous open questions related to filament
interactions and non-equilibrium solutions; see, e.g.,
\cite{kwiecinski2018dynamics} for some recent results on two-filament
interactions. But even for equlibrium states, a whole zoo of other
\textquotedblleft exotic\textquotedblright\ (non-helical)\ relative equlibria
exist, as illustrated in Figure \ref{fig:zoo} in connection to the gradient
flow of Eq.~(\ref{ener}). In this figure, we observe different types of
energetically stable equilibria resulting from initial conditions with
different combinations of $(K,q)$, for a few distinct values of $D$. The
helical states we considered in this paper only scratch the surface of what is
possible and provide a sense of the wealth of associated possibilities.
Understanding the emergence/bifurcations of such states, but also the energy
landscape and dynamics associated with them provide, in our view, a
fascinating potential for further exploration.

\bibliographystyle{elsarticle-num}
\bibliography{fila_new}

\begin{thebibliography}{10}
\expandafter\ifx\csname url\endcsname\relax
  \def\url#1{\texttt{#1}}\fi
\expandafter\ifx\csname urlprefix\endcsname\relax\def\urlprefix{URL }\fi
\expandafter\ifx\csname href\endcsname\relax
  \def\href#1#2{#2} \def\path#1{#1}\fi

\bibitem{Pismen}
L.~Pismen, Vortices in Nonlinear Fields, Oxford University Press, Clarendon,
  UK, 1999.

\bibitem{Donnelly}
R.~Donnelly, Quantized Vortices in Helium II, Cambridge University Press,
  Cambridge, UK, 1991.

\bibitem{Blatter1994}
G.~Blatter, M.~V. Feigel'man, V.~B. Geshkenbein, A.~I. Larkin, V.~M. Vinokur,
  Vortices in high-temperature superconductors, Rev. Mod. Phys. 66 (1994)
  1125--1388.

\bibitem{Saffman}
P.~Saffman, Vortex Dynamics, Cambridge University Press, Cambridge, UK, 1992.

\bibitem{becbook1}
L.~Pitaevskii, S.~Stringari, {B}ose--{E}instein Condensation, Oxford University
  Press, Oxford, UK, 2003.

\bibitem{becbook2}
C.~Pethick, H.~Smith, {B}ose--{E}instein Condensation in Dilute Gases,
  Cambridge University Press, Cambridge, UK, 2002.

\bibitem{siambook}
P.~Kevrekidis, D.~Frantzeskakis, R.~Carretero-González, The Defocusing
  Nonlinear Schrodinger Equation, Society for Industrial and Applied
  Mathematics, Philadelphia, PA, 2015.

\bibitem{fetter2}
A.~L. Fetter, Rotating trapped bose-einstein condensates, Rev. Mod. Phys. 81
  (2009) 647--691.

\bibitem{Alexander2001}
A.~L. Fetter, A.~A. Svidzinsky, Vortices in a trapped dilute bose-einstein
  condensate, Journal of Physics: Condensed Matter 13~(12) (2001) R135.

\bibitem{mplb04}
P.~Kevrekidis, R.~Carretero-Gonz{\'a}lez, D.~Frantzeskakis, I.~Kevrekidis,
  Vortices in bose-einstein condensates: some recent developments, Modern
  Physics Letters B 18~(30) (2004) 1481--1505.

\bibitem{Komineas2007}
S.~Komineas, Vortex rings and solitary waves in trapped bose--einstein
  condensates, The European Physical Journal Special Topics 147~(1) (2007)
  133--152.

\bibitem{White2013}
A.~C. White, B.~P. Anderson, V.~S. Bagnato, Vortices and turbulence in trapped
  atomic condensates, Proceedings of the National Academy of Sciences (2014).

\bibitem{TSATSOS20161}
M.~C. Tsatsos, P.~E. Tavares, A.~Cidrim, A.~R. Fritsch, M.~A. Caracanhas,
  F.~E.~A. dos Santos, C.~F. Barenghi, V.~S. Bagnato, Quantum turbulence in
  trapped atomic bose–einstein condensates, Physics Reports 622 (2016) 1 --
  52, quantum turbulence in trapped atomic Bose–Einstein condensates.

\bibitem{RSB1999}
R.~Ricca, D.~Samuels, C.~Barenghi, Evolution of vortex knots, Journal of Fluid
  Mechanics 391 (1999) 29–44.

\bibitem{pre09}
F.~Maggioni, S.~Alamri, C.~F. Barenghi, R.~L. Ricca, Velocity, energy, and
  helicity of vortex knots and unknots, Phys. Rev. E 82 (2010) 026309.

\bibitem{pre12}
F.~Maggioni, S.~Alamri, C.~F. Barenghi, R.~L. Ricca, Velocity, energy, and
  helicity of vortex knots and unknots, Phys. Rev. E 82 (2010) 026309.

\bibitem{nature13}
D.~Kleckner, W.~Irvine, Creation and dynamics of knotted vortices, Nature Phys.
  9 (2013) 253--258.

\bibitem{Proment_2014}
D.~Proment, M.~Onorato, C.~F. Barenghi, Torus quantum vortex knots in the
  gross-pitaevskii model for bose-einstein condensates, Journal of Physics:
  Conference Series 544 (2014) 012022.

\bibitem{brachet16}
P.~Clark~di Leoni, P.~D. Mininni, M.~E. Brachet, Helicity, topology, and kelvin
  waves in reconnecting quantum knots, Phys. Rev. A 94 (2016) 043605.

\bibitem{nature16}
D.~Kleckner, W.~Kauffman, L.H.~Irvine, How superfluid vortex knots untie,
  Nature Phys. 12 (2016) 650--655.

\bibitem{Ruban2018a}
V.~P. Ruban, Long-lived quantum vortex knots, JETP Letters 107~(5) (2018)
  307--310.

\bibitem{Ruban2018b}
V.~P. Ruban, Quasi-stable configurations of torus vortex knots and links,
  Journal of Experimental and Theoretical Physics 127~(3) (2018) 581--586.

\bibitem{rubanus}
C.~Ticknor, V.~P. Ruban, P.~G. Kevrekidis, Quasistable quantum vortex knots and
  links in anisotropic harmonically trapped bose-einstein condensates, Phys.
  Rev. A 99 (2019) 063604.

\bibitem{nature16b}
D.~Hall, M.~Ray, K.~Tiurev, R.~E., A.~Gheorghe, M.~M{\"o}tt{\"o}nen, Tying
  quantum knots, Nature Phys. 12 (2016) 478--483.

\bibitem{Leeeaao3820}
W.~Lee, A.~H. Gheorghe, K.~Tiurev, T.~Ollikainen, M.~M{\"o}tt{\"o}nen, D.~S.
  Hall, Synthetic electromagnetic knot in a three-dimensional skyrmion, Science
  Advances 4~(3) (2018).

\bibitem{scirep12}
A.~Desyatnikov, D.~Buccoliero, M.~Dennis, Y.~Kivshar, Spontaneous knotting of
  self-trapped waves, Sci. Rep. 2:771 (2012) 1--7.

\bibitem{Shimokawa20906}
K.~Shimokawa, K.~Ishihara, I.~Grainge, D.~J. Sherratt, M.~Vazquez,
  Ftsk-dependent xercd-dif recombination unlinks replication catenanes in a
  stepwise manner, Proceedings of the National Academy of Sciences 110~(52)
  (2013) 20906--20911.

\bibitem{nature13b}
J.~Cirtain, L.~Golub, A.~Winebarger, B.~De~Pontieu, K.~Kobayashi, R.~Moore,
  R.~Walsh, K.~Korreck, M.~Weber, P.~McCauley, T.~A., S.~Kuzin, D.~C.E., Energy
  release in the solar corona from spatially resolved magnetic braids, Nature
  493 (2013) 501--503.

\bibitem{lim1997role}
T.~Lim, On the role of kelvin-helmholtz-like instability in the formation of
  turbulent vortex rings, Fluid dynamics research 21~(1) (1997) 47--56.

\bibitem{caplan2014scattering}
R.~Caplan, J.~Talley, R.~Carretero-Gonz{\'a}lez, P.~Kevrekidis, Scattering and
  leapfrogging of vortex rings in a superfluid, Physics of Fluids 26~(9) (2014)
  097101.

\bibitem{baggaley2011spectrum}
A.~W. Baggaley, C.~F. Barenghi, Spectrum of turbulent kelvin-waves cascade in
  superfluid helium, Physical Review B 83~(13) (2011) 134509.

\bibitem{okulov2010maximum}
V.~Okulov, J.~N. S{\o}rensen, Maximum efficiency of wind turbine rotors using
  joukowsky and betz approaches, Journal of Fluid Mechanics 649 (2010)
  497--508.

\bibitem{okulov2004stability}
V.~Okulov, On the stability of multiple helical vortices, Journal of Fluid
  Mechanics 521 (2004) 319--342.

\bibitem{quaranta2015long}
H.~U. Quaranta, H.~Bolnot, T.~Leweke, Long-wave instability of a helical
  vortex, Journal of Fluid Mechanics 780 (2015) 687--716.

\bibitem{leweke2014long}
T.~Leweke, H.~Quaranta, H.~Bolnot, F.~Blanco-Rodr{\'\i}guez, S.~Le~Diz{\`e}s,
  Long-and short-wave instabilities in helical vortices, in: Journal of
  Physics: Conference Series, Vol. 524, IOP Publishing, 2014, p. 012154.

\bibitem{klein1995simplified}
R.~Klein, A.~J. Majda, K.~Damodaran, Simplified equations for the interaction
  of nearly parallel vortex filaments, Journal of Fluid Mechanics 288 (1995)
  201--248.

\bibitem{lions2000equilibrium}
P.-L. Lions, A.~Majda, Equilibrium statistical theory for nearly parallel
  vortex filaments, Communications on Pure and Applied Mathematics: A Journal
  Issued by the Courant Institute of Mathematical Sciences 53~(1) (2000)
  76--142.

\bibitem{contreras2016nearly}
A.~Contreras, R.~L. Jerrard, Nearly parallel vortex filaments in the 3d
  ginzburg-landau equations, arXiv preprint arXiv:1606.00732 (2016).

\bibitem{barry2012relative}
A.~Barry, G.~Hall, C.~Wayne, Relative equilibria of the (1+ n)-vortex problem,
  Journal of nonlinear science 22~(1) (2012) 63--83.

\bibitem{newton2001n}
P.~Newton, The N-vortex problem: analytical techniques, Vol. 145, Springer,
  2001.

\bibitem{aref2003vortex}
H.~Aref, P.~Newton, M.~Stremler, T.~Tokieda, D.~Vainchtein, Vortex crystals,
  Advances in applied Mechanics 39 (2003) 1--79.

\bibitem{palmore1982relative}
J.~Palmore, Relative equilibria of vortices in two dimensions, Proceedings of
  the National Academy of Sciences 79~(2) (1982) 716--718.

\bibitem{DGPE}
D.~Yan, R.~Carretero-Gonz\'alez, D.~J. Frantzeskakis, P.~G. Kevrekidis, N.~P.
  Proukakis, D.~Spirn, Exploring vortex dynamics in the presence of
  dissipation: Analytical and numerical results, Phys. Rev. A 89 (2014) 043613.

\bibitem{zueva}
T.~I. Zueva, Dissipative motion of vortices in spatially inhomogeneous
  bose-einstein condensates, Low Temperature Physics 45~(1) (2019) 67--77.

\bibitem{yongil}
G.~Moon, W.~J. Kwon, H.~Lee, Y.-i. Shin, Thermal friction on quantum vortices
  in a bose-einstein condensate, Phys. Rev. A 92 (2015) 051601.

\bibitem{ruste}
J.~Ruostekoski, Z.~Dutton, Dynamical and energetic instabilities in
  multicomponent bose-einstein condensates in optical lattices, Phys. Rev. A 76
  (2007) 063607.

\bibitem{chen2013collective}
Y.~Chen, T.~Kolokolnikov, D.~Zhirov, Collective behaviour of large number of
  vortices in the plane, Proceedings of the Royal Society A: Mathematical,
  Physical and Engineering Sciences 469~(2156) (2013) 20130085.

\bibitem{kolokolnikov2014tale}
T.~Kolokolnikov, P.~Kevrekidis, R.~Carretero-Gonz{\'a}lez, A tale of two
  distributions: from few to many vortices in quasi-two-dimensional
  bose--einstein condensates, Proceedings of the Royal Society A: Mathematical,
  Physical and Engineering Sciences 470~(2168) (2014) 20140048.

\bibitem{Kolokolnikov:2010}
T.~Kolokolnikov, H.~Sun, D.~Uminsky, A.~L. Bertozzi, Stability of ring patterns
  arising from two-dimensional particle interactions, Phys. Rev. E Rapid. Comm.
  84 (2011) 015203.

\bibitem{havelock}
T.~Havelock, Lii. the stability of motion of rectilinear vortices in ring
  formation, The London, Edinburgh, and Dublin Philosophical Magazine and
  Journal of Science 11~(70) (1931) 617--633.

\bibitem{FBD2010}
C.~J. Foster, P.~B. Blakie, M.~J. Davis, Vortex pairing in two-dimensional bose
  gases, Phys. Rev. A 81 (2010) 023623.

\bibitem{top-2017}
R.~N. Bisset, S.~Serafini, E.~Iseni, M.~Barbiero, T.~Bienaim\'e, G.~Lamporesi,
  G.~Ferrari, F.~Dalfovo, Observation of a spinning top in a bose-einstein
  condensate, Phys. Rev. A 96 (2017) 053605.

\bibitem{Villois2016}
A.~Villois, G.~Krstulovic, D.~Proment, H.~Salman, A vortex filament tracking
  method for the gross{\textendash}pitaevskii model of a superfluid, Journal of
  Physics A: Mathematical and Theoretical 49~(41) (2016) 415502.

\bibitem{kwiecinski2018dynamics}
J.~A. Kwiecinski, R.~A. Van~Gorder, Dynamics of nearly parallel interacting
  vortex filaments, Journal of Fluid Mechanics 835 (2018) 575--623.

\end{thebibliography}

\end{document}